\documentclass[usenatbib]{aa}     

\usepackage{graphicx}
\usepackage{txfonts}
\usepackage{longtable}
\usepackage{multirow}
\def\crs{\cr\noalign{\smallskip}}

\def\EQM#1{\vcenter{\normalbaselines{\mathsurround=0pt}
    \ialign{${\displaystyle ##}$\hfil&&\ ${\displaystyle ##}$\hfil\crcr
    \mathstrut\crcr\noalign{\kern-\baselineskip}
    \noalign{\smallskip}
    #1\crcr\mathstrut\crcr\noalign{\kern-\baselineskip}}}}
\def\e{{\rm e}}
\def\i{{\rm i}}

\newcommand\tableA{
\begin{table} 
      \caption[]{Summary of procedure parameters.}
\renewcommand{\arraystretch}{1}
\begin{center}
\small
\begin{tabular}{ccc}
\hline
   & Meaning & Venue \\
\hline
\crs
 \multirow{ 2}{*}{$M_n$} & Degree of the polynomial &  \multirow{ 2}{*}{(\ref{for nu Cheb})} \\
                         & approximating the $n$th fundamental frequency &  \vspace{2mm}\crs
 \multirow{ 2}{*}{$K_n$} & Maximum index number &  \multirow{ 2}{*}{(\ref{for frequency index set})} \\
                         & associated with the $n$th fundamental frequency \vspace{2mm}\crs
 \multirow{ 2}{*}{$L_\mathbf{k}$} & Degree of the polynomial &  \multirow{ 2}{*}{(\ref{for Cheb app of amplitude})} \\
                         & approximating the amplitude with index $\mathbf{k}$ \vspace{2mm}\crs
 $J$ & Maximum number of representation terms & \crs\\
 $\delta$ & Absolute truncation error & (\ref{for absolute error}) \vspace{2mm}\crs
 $\delta_r$ & Relative truncation error & (\ref{for relative error}) \vspace{2mm}\crs
  \hline
\label{tab procedure parameters}
\end{tabular}
\end{center}
\renewcommand{\arraystretch}{1}
\end{table}
}

\newcommand\tableB{
\begin{table} 
      \caption[]{The coefficients of the Chebyshev expansion $\sum_{m=0}^{9}c_m
T_m(x(t))$ approximating $\nu(t)$, the changing fundamental frequency of the
dissipated solution specified by (\ref{for D equation}), (\ref{for
D eqnparameter}) and (\ref{for D initialcondition}).}
\renewcommand{\arraystretch}{1.1}
\begin{center}
\begin{tabular}{ccc}
\hline
  $m$ & $c_m$ \\
\hline
  0 &  $+ 1.450265\times 10^{ 0}$ \crs
  1 &  $- 1.032502\times 10^{-1}$ \crs
  2 &  $- 7.924442\times 10^{-4}$ \crs
  3 &  $- 1.163806\times 10^{-4}$ \crs
  4 &  $- 7.914948\times 10^{-6}$ \crs
  5 &  $- 6.369786\times 10^{-7}$ \crs
  6 &  $- 5.975314\times 10^{-8}$ \crs
  7 &  $- 3.609371\times 10^{-9}$ \crs
  8 &  $+ 1.771340\times 10^{-9}$ \crs
  9 &  $+ 8.159405\times 10^{-9}$ \crs
  \hline
\label{tab D FFdata}
\end{tabular}
\end{center}
\renewcommand{\arraystretch}{1}
\end{table}
}

\newcommand\tableC{
\begin{table} 
      \caption[]{The leading 10 terms of (\ref{37 term}), a representation of the dissipated solution $z(t)$ specified by (\ref{for D equation}), (\ref{for
D eqnparameter}) and (\ref{for D initialcondition}), which is obtained with $\{M,K,L,\delta_r\}=\{9,10,9,10^{-5}\}$}.
\renewcommand{\arraystretch}{1.1}
\begin{center}
\begin{tabular}{ccccc}
\hline
j & $l(j)$ & $k(j)$ & $Arg(a_j)$ & $|a_j|$  \\
\hline
  1 & 0 & 1 & $ -1.431544 $ & $  1.378074489 $ \crs
  2 & 0 & 2 & $  0.278103 $ & $  0.622837454 $ \crs
  3 & 0 & 3 & $  1.987348 $ & $  0.159698128 $ \crs
  4 & 1 & 1 & $  1.748566 $ & $  0.116187680 $ \crs
  5 & 0 & 4 & $ -2.587002 $ & $  0.032622276 $ \crs
  6 & 0 &-1 & $ -1.709654 $ & $  0.017747854 $ \crs
  7 & 1 & 2 & $  0.133848 $ & $  0.028210547 $ \crs
  8 & 1 & 3 & $  1.931736 $ & $  0.027791194 $ \crs
  9 & 0 & 5 & $ -0.878547 $ & $  0.005904572 $ \crs
 10 & 1 & 4 & $ -2.628911 $ & $  0.009823726 $ \crs
\hline
\label{tab D termdata}
\end{tabular}
\end{center}
\renewcommand{\arraystretch}{1}
   \end{table}
}

\newcommand\tableD{
\begin{table} 
\caption[]{The coefficients of Chebyshev expansion $\sum_m c_m
T_m$ approximating the major fundamental frequencies $(g_1,...,g_8,s_1,..,s_8)$ (arcsec\ yr$^{-1}$) of the solar system over the time interval from -35 Myr to 5 Myr with origin at J2000.  Also listed are the used constant libration frequencies  $r_1$ and $r_2$ (arcsec\ yr$^{-1}$).} \label{tab FFChC}
\begin{center}
\scriptsize
\begin{tabular}{|c|cr||c|cr|}
\hline
  & $m$& $c_m $ &  & $m$ & $c_m$ \\
 \hline
 \multirow{ 16}{*}{$g_1$} & $ 0 $ & $   5.59436858 $ &  \multirow{ 16}{*}{$s_1$} & $ 0 $ & $  -5.61412432 $ \\
  & $  1 $ & $  -0.02916946 $  &  & $  1 $ & $   0.01897975 $ \\
  & $  2 $ & $   0.00259949 $  &  & $  2 $ & $   0.00616405 $ \\
  & $  3 $ & $  -0.00858191 $  &  & $  3 $ & $  -0.00967633 $ \\
  & $  4 $ & $  -0.00538351 $  &  & $  4 $ & $  -0.00048289 $ \\
  & $  5 $ & $  -0.00216691 $  &  & $  5 $ & $  -0.00405450 $ \\
  & $  6 $ & $  -0.00311006 $  &  & $  6 $ & $  -0.00319098 $ \\
  & $  7 $ & $   0.00137912 $  &  & $  7 $ & $   0.00116236 $ \\
  & $  8 $ & $  -0.00225709 $  &  & $  8 $ & $  -0.00048593 $ \\
  & $  9 $ & $   0.00318074 $  &  & $  9 $ & $   0.00207233 $ \\
  & $ 10 $ & $   0.00431717 $  &  & $ 10 $ & $   0.00194425 $ \\
  & $ 11 $ & $  -0.00288184 $  &  & $ 11 $ & $  -0.00103824 $ \\
  & $ 12 $ & $  -0.00189708 $  &  & $ 12 $ & $  -0.00117501 $ \\
  & $ 13 $ & $   0.00127879 $  &  & $ 13 $ & $   0.00006311 $ \\
  & $ 14 $ & $   0.00065723 $  &  & $ 14 $ & $   0.00030964 $ \\
  & $ 15 $ & $  -0.00017362 $  &  & $ 15 $ & $  -0.00037496 $ \\
 \hline
 \multirow{ 16}{*}{$g_2$} & $ 0 $ & $   7.45660678 $ &  \multirow{ 16}{*}{$s_2$} & $ 0 $ & $  -7.07313208 $ \\
  & $  1 $ & $  -0.00276205 $  &  & $  1 $ & $   0.04218433 $ \\
  & $  2 $ & $   0.00177713 $  &  & $  2 $ & $  -0.00001315 $ \\
  & $  3 $ & $   0.00056064 $  &  & $  3 $ & $  -0.00225169 $ \\
  & $  4 $ & $  -0.00002096 $  &  & $  4 $ & $   0.00386256 $ \\
  & $  5 $ & $   0.00090782 $  &  & $  5 $ & $  -0.00280878 $ \\
  & $  6 $ & $   0.00096229 $  &  & $  6 $ & $   0.00247914 $ \\
  & $  7 $ & $  -0.00024908 $  &  & $  7 $ & $   0.00109449 $ \\
  & $  8 $ & $   0.00042855 $  &  & $  8 $ & $  -0.00064733 $ \\
  & $  9 $ & $  -0.00071228 $  &  & $  9 $ & $   0.00229722 $ \\
  & $ 10 $ & $  -0.00091971 $  &  & $ 10 $ & $  -0.00000725 $ \\
  & $ 11 $ & $   0.00061668 $  &  & $ 11 $ & $  -0.00193901 $ \\
  & $ 12 $ & $   0.00047006 $  &  & $ 12 $ & $   0.00088350 $ \\
  & $ 13 $ & $  -0.00024111 $  &  & $ 13 $ & $   0.00053257 $ \\
  & $ 14 $ & $  -0.00015053 $  &  & $ 14 $ & $   0.00024897 $ \\
  & $ 15 $ & $   0.00007778 $  &  & $ 15 $ & $   0.00024128 $ \\
 \hline
 \multirow{ 16}{*}{$g_3$} & $ 0 $ & $  17.36445990 $ &  \multirow{ 16}{*}{$s_3$} & $ 0 $ & $ -18.84810087 $ \\
  & $  1 $ & $   0.01585020 $  &  & $  1 $ & $  -0.00165582 $ \\
  & $  2 $ & $  -0.00496995 $  &  & $  2 $ & $   0.00007105 $ \\
  & $  3 $ & $   0.00540088 $  &  & $  3 $ & $  -0.00140634 $ \\
  & $  4 $ & $   0.00496032 $  &  & $  4 $ & $   0.00050744 $ \\
  & $  5 $ & $  -0.00226103 $  &  & $  5 $ & $   0.00039039 $ \\
  & $  6 $ & $  -0.00004476 $  &  & $  6 $ & $   0.00135948 $ \\
  & $  7 $ & $   0.00006943 $  &  & $  7 $ & $   0.00065306 $ \\
  & $  8 $ & $   0.00007190 $  &  & $  8 $ & $  -0.00041702 $ \\
  & $  9 $ & $  -0.00037475 $  &  & $  9 $ & $  -0.00026792 $ \\
  & $ 10 $ & $  -0.00059252 $  &  & $ 10 $ & $  -0.00018687 $ \\
  & $ 11 $ & $   0.00025574 $  &  & $ 11 $ & $   0.00004492 $ \\
  & $ 12 $ & $   0.00035119 $  &  & $ 12 $ & $   0.00013288 $ \\
  & $ 13 $ & $  -0.00020600 $  &  & $ 13 $ & $   0.00001794 $ \\
  & $ 14 $ & $  -0.00000042 $  &  & $ 14 $ & $  -0.00005746 $ \\
  & $ 15 $ & $   0.00011279 $  & & & \\
 \hline
 \multirow{ 16}{*}{$g_4$} & $ 0 $ & $  17.91086281 $ &  \multirow{ 16}{*}{$s_4$} & $ 0 $ & $ -17.75496646 $ \\
  & $  1 $ & $   0.02044603 $  &  & $  1 $ & $   0.00670633 $ \\
  & $  2 $ & $  -0.00851446 $  &  & $  2 $ & $  -0.00717162 $ \\
  & $  3 $ & $   0.00812180 $  &  & $  3 $ & $   0.00313019 $ \\
  & $  4 $ & $   0.00866961 $  &  & $  4 $ & $   0.00807398 $ \\
  & $  5 $ & $  -0.00482118 $  &  & $  5 $ & $  -0.00498282 $ \\
  & $  6 $ & $  -0.00137335 $  &  & $  6 $ & $  -0.00116638 $ \\
  & $  7 $ & $   0.00011721 $  &  & $  7 $ & $   0.00058378 $ \\
  & $  8 $ & $   0.00083465 $  &  & $  8 $ & $   0.00021726 $ \\
  & $  9 $ & $  -0.00041179 $  &  & $  9 $ & $   0.00016650 $ \\
  & $ 10 $ & $  -0.00109334 $  &  & $ 10 $ & $  -0.00060643 $ \\
  & $ 11 $ & $   0.00069484 $  &  & $ 11 $ & $   0.00014173 $ \\
  & $ 12 $ & $   0.00045529 $  &  & $ 12 $ & $   0.00039520 $ \\
  & $ 13 $ & $  -0.00029042 $  &  & $ 13 $ & $  -0.00024600 $ \\
  & $ 14 $ & $  -0.00014170 $  &  & $ 14 $ & $  -0.00006598 $ \\
  & $ 15 $ & $   0.00008189 $  &  & $ 15 $ & $   0.00012665 $ \\
 \hline
 \multirow{  1}{*}{$g_5$} & $ 0 $ & $   4.25745185 $ &  \multirow{  1}{*}{$s_5$} & $ 0 $ & $   0.00000015 $ \\
 \hline
 \multirow{  5}{*}{$g_6$} & $ 0 $ & $  28.24498422 $ &  \multirow{  5}{*}{$s_6$} & $ 0 $ & $ -26.34785292 $ \\
  & $  1 $ & $   0.00010582 $  &  & $  1 $ & $   0.00000053 $ \\
  & $  2 $ & $   0.00011695 $  &  & $  2 $ & $  -0.00000514 $ \\
  & $  3 $ & $  -0.00002698 $  & & & \\
  & $  4 $ & $  -0.00001157 $  & & & \\
 \hline
 \multirow{  2}{*}{$g_7$} & $ 0 $ & $   3.08795246 $ &  \multirow{  2}{*}{$s_7$} & $ 0 $ & $  -2.99252583 $ \\
  & $  1 $ & $  -0.00000017 $  & & & \\
 \hline
 \multirow{  2}{*}{$g_8$} & $ 0 $ & $   0.67302182  $ &  \multirow{  2}{*}{$s_8$} & $ 0 $ & $  -0.69173649 $ \\
  & &  & & $  1  $ & $   0.00000001 $ \\
\hline
\hline
$r_1$ & $ 0 $ & $ 0.251085\ \ \ \ $ &  $r_2$ & $ 0 $ & $ 0.117222\ \ \ \ $ \\
\hline
\end{tabular}
\end{center}
\end{table}
}

\newcommand\tableE{
\begin{table*} 
\caption[]{The leading 40 terms of the 100-term representation $z_3(t)$, as  expressed in (\ref{z zata 100}). \label{tab MainFrequencyz3}}
\begin{center}
\begin{tabular}{ccrr}
  \hline
 No. & $\langle \mathbf{k},\mathbf{f} \rangle$ & $\mathrm{Abs}(a_\mathbf{k}) \times 10^6$ & $\mathrm{Arg}(a_\mathbf{k})$ (degree)  \\
  \hline
   1            & $   g_5                    $ & $  18984 $ & $  -68.812 $\crs
   2            & $   g_2                    $ & $  16088 $ & $   95.535 $\crs
   3            & $   g_4                    $ & $  13041 $ & $   17.345 $\crs
   4            & $   g_3                    $ & $   9042 $ & $  -54.984 $\crs
   5            & $   g_1                    $ & $   4314 $ & $ -175.273 $\crs
   6            & $   g_4- r_1               $ & $   2583 $ & $  -92.093 $\crs
   7            & $   g_3- r_1               $ & $   2415 $ & $   15.391 $\crs
   8            & $   g_3+ r_1+ r_2          $ & $   2377 $ & $ -136.939 $\crs
   9            & $   g_4+ s_3- s_4          $ & $   1934 $ & $ -128.504 $\crs
  10            & $   g_6                    $ & $   1498 $ & $  160.676 $\crs
  11            & $  2g_1- g_5               $ & $   1393 $ & $   82.417 $\crs
  12            & $   g_3+ g_4- g_6          $ & $   1372 $ & $  178.968 $\crs
  13            & $   g_1- r_2               $ & $   1298 $ & $  140.905 $\crs
  14            & $   g_3- s_3+ s_4          $ & $   1282 $ & $  -87.475 $\crs
  15            & $   g_2- r_2               $ & $   1156 $ & $ -127.879 $\crs
  16            & $   g_4-2r_1               $ & $   1153 $ & $   15.739 $\crs
  17            & $   g_1+ r_2               $ & $   1085 $ & $   49.430 $\crs
  18            & $   g_2- g_3+ g_6- r_2     $ & $   1028 $ & $  -40.633 $\crs
  19            & $   g_4- r_2               $ & $    946 $ & $  149.088 $\crs
  20            & $   g_2+ r_2               $ & $    942 $ & $  123.828 $\crs
  21            & $ - g_1+2g_4- g_5+ s_3     $ & $    916 $ & $   82.353 $\crs
  22            & $ - g_3+2g_4-2r_1          $ & $    903 $ & $   63.075 $\crs
  23            & $   g_3+ s_3- s_4          $ & $    824 $ & $  150.057 $\crs
  24            & $   g_3+2r_1               $ & $    816 $ & $  145.441 $\crs
  25            & $   g_4+ r_1               $ & $    806 $ & $  165.483 $\crs
  26            & $   g_4+ s_3- s_4- r_1     $ & $    756 $ & $  113.428 $\crs
  27            & $   g_1- s_3+ s_4+ r_2     $ & $    712 $ & $ -121.353 $\crs
  28            & $   g_4- s_3+ s_4          $ & $    697 $ & $  -14.943 $\crs
  29            & $  2g_3- g_4+2r_2          $ & $    605 $ & $  -21.586 $\crs
  30            & $   g_7                    $ & $    577 $ & $ -146.073 $\crs
  31            & $   g_4+ r_2               $ & $    573 $ & $   52.634 $\crs
  32            & $   g_3+ r_2               $ & $    504 $ & $  -21.947 $\crs
  33            & $   g_1+ g_5- g_7+ r_1     $ & $    432 $ & $  152.449 $\crs
  34            & $ - g_1+ g_2+ g_5- r_2     $ & $    383 $ & $  -28.070 $\crs
  35            & $   g_3+ r_1               $ & $    275 $ & $   -4.253 $\crs
  36            & $   g_3- r_2               $ & $    144 $ & $   89.319 $\crs
  37            & $   g_2- r_1- r_2          $ & $    121 $ & $  -36.562 $\crs
  38            & $   g_2- r_1+ r_2          $ & $    104 $ & $ -105.175 $\crs
  39            & $ - g_1+ g_4+ g_5          $ & $     92 $ & $ -173.117 $\crs
  40            & $   g_1- r_1               $ & $     87 $ & $   99.123 $\crs
\hline
\end{tabular}
\end{center}
\end{table*}
}

\newcommand\tableF{
\begin{table*} 
\caption[]{The leading 40 terms of the 100-term representation $\zeta_3(t)$, as expressed in (\ref{z zata 100}). \label{tab MainFrequencyzeta3}}
\begin{center}
\begin{tabular}{ccrr}
\hline
  No. & $ \langle \mathbf{k},\mathbf{f} \rangle$ & $\mathrm{Abs}(b_\mathbf{k}) \times 10^6$ & $\mathrm{Arg}(b_\mathbf{k})$ (degree)  \\
\hline
   1            & $   s_5                    $ & $  13774 $ & $  107.587 $\crs
   2            & $   s_3                    $ & $   8666 $ & $  -62.318 $\crs
   3            & $   s_4                    $ & $   4647 $ & $   96.756 $\crs
   4            & $   s_1                    $ & $   4085 $ & $   27.817 $\crs
   5            & $   s_2                    $ & $   3312 $ & $   80.364 $\crs
   6            & $   g_3- g_4+ s_4          $ & $   2745 $ & $ -167.132 $\crs
   7            & $   s_2+2r_2               $ & $   2041 $ & $  -44.496 $\crs
   8            & $   s_2+ r_2               $ & $   1543 $ & $  125.410 $\crs
   9            & $   g_3- g_4+ s_3          $ & $   1530 $ & $ -137.669 $\crs
  10            & $   s_1+ s_3- s_4- r_2     $ & $   1469 $ & $   91.798 $\crs
  11            & $   s_2- r_2               $ & $   1450 $ & $   23.071 $\crs
  12            & $   s_1+ r_2               $ & $   1417 $ & $ -116.942 $\crs
  13            & $   s_6                    $ & $   1333 $ & $  110.029 $\crs
  14            & $   s_7                    $ & $    889 $ & $    9.186 $\crs
  15            & $   s_2+ r_1               $ & $    646 $ & $  -50.277 $\crs
  16            & $   s_8                    $ & $    641 $ & $   26.053 $\crs
  17            & $   s_3- r_1               $ & $    613 $ & $ -178.905 $\crs
  18            & $   s_1+ s_3- s_4          $ & $    532 $ & $ -177.568 $\crs
  19            & $   s_1-2r_2               $ & $    518 $ & $  101.955 $\crs
  20            & $   s_3- r_2               $ & $    484 $ & $  106.762 $\crs
  21            & $   g_3- g_4+ s_3+ r_1     $ & $    481 $ & $  -21.125 $\crs
  22            & $   g_3- g_4+ s_2+ r_1     $ & $    445 $ & $   -5.831 $\crs
  23            & $   s_2- s_3+ s_4          $ & $    364 $ & $   72.700 $\crs
  24            & $   s_2- s_3+ s_4- r_2     $ & $    344 $ & $   31.282 $\crs
  25            & $   s_2- r_1               $ & $    341 $ & $   20.734 $\crs
  26            & $   s_2-2r_2               $ & $    320 $ & $  -51.578 $\crs
  27            & $   g_3- g_4+ s_1          $ & $    315 $ & $ -118.127 $\crs
  28            & $   g_3- s_1+ s_6+ s_7- s_8$ & $    304 $ & $  172.667 $\crs
  29            & $   s_1- r_1               $ & $    293 $ & $  149.314 $\crs
  30            & $ - s_4+ s_6- s_7+2s_8     $ & $    293 $ & $  -76.858 $\crs
  31            & $   s_1+2r_1- r_2          $ & $    292 $ & $  134.167 $\crs
  32            & $ - g_3+ s_5+ s_8+ r_1     $ & $    285 $ & $ -170.886 $\crs
  33            & $ - g_3+ g_4+ s_2          $ & $    268 $ & $   47.758 $\crs
  34            & $ - g_3+ g_4+ s_2- r_2     $ & $    258 $ & $  -27.836 $\crs
  35            & $ - g_3+ g_4+ s_4          $ & $    244 $ & $  -18.592 $\crs
  36            & $   g_4+ s_6- s_8+ r_1     $ & $    230 $ & $  109.121 $\crs
  37            & $   g_3- g_4+ s_4- r_1     $ & $    221 $ & $  -91.732 $\crs
  38            & $   s_2+ s_5- s_8-2r_1     $ & $    217 $ & $  173.446 $\crs
  39            & $ - g_3+ g_4+ s_2- r_1     $ & $    198 $ & $   24.257 $\crs
  40            & $   s_1+2r_1               $ & $     78 $ & $   77.412 $\crs
 \hline
\end{tabular}
\end{center}
\end{table*}
}

\newcommand\figureA{
   \begin{figure} 
   \centering
   \includegraphics[width=100mm]{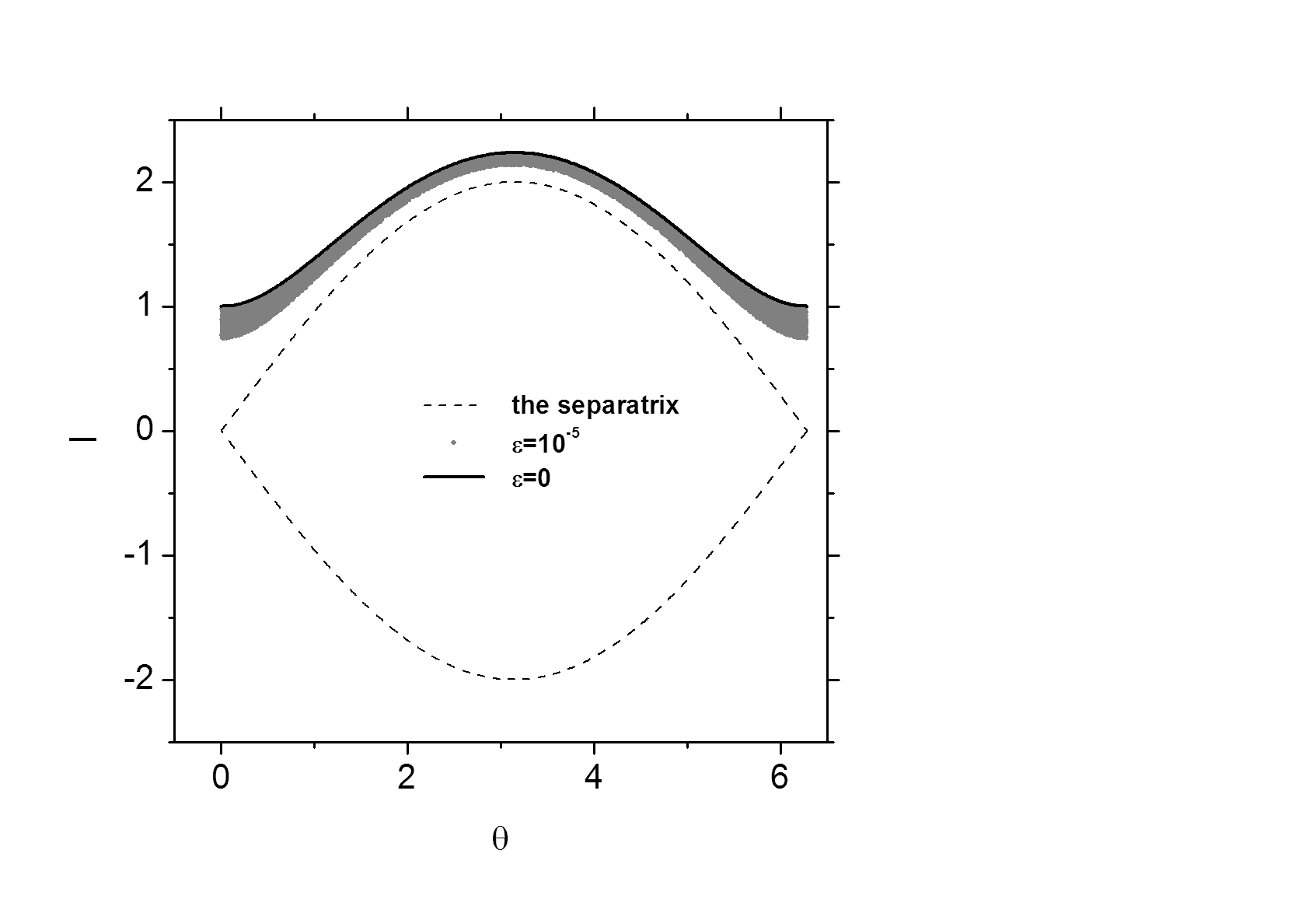}
      \caption{Phase trajectory of the dissipated solution specified by
(\ref{for D equation}), (\ref{for D eqnparameter}) and (\ref{for D
initialcondition}). Also shown is the phase orbit of the unperturbed pendulum system, namely (\ref{for D equation}) with $\varepsilon=0$, which passes through the initial phase point of the dissipated solution. The dashed line depicts the spepratrix of the unperturbed pendulum.} \label{fig D orbit}
   \end{figure}
}

\newcommand\figureB{
   \begin{figure} 
   \centering
      \includegraphics[width=100mm]{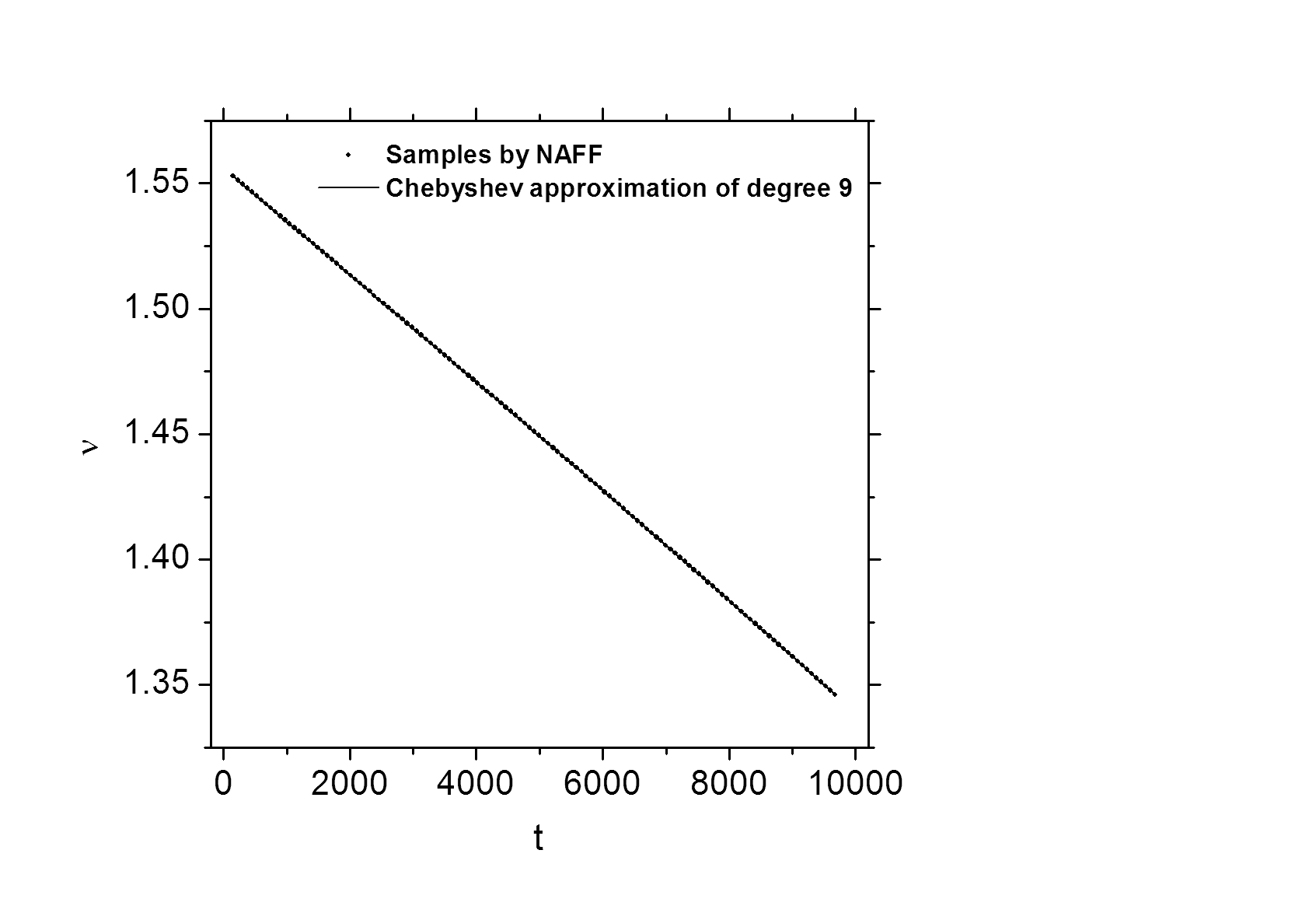}
      \caption{The changing fundamental frequency $\nu(t)$ of the
dissipated solution specified by (\ref{for D equation}), (\ref{for
D eqnparameter}) and (\ref{for D initialcondition}). Also shown is a Chebyshev approximation of $\nu(t)$.} \label{fig D tvsff}
   \end{figure}
}

\newcommand\figureC{
\begin{figure} 
   \centering
         \includegraphics[width=100mm]{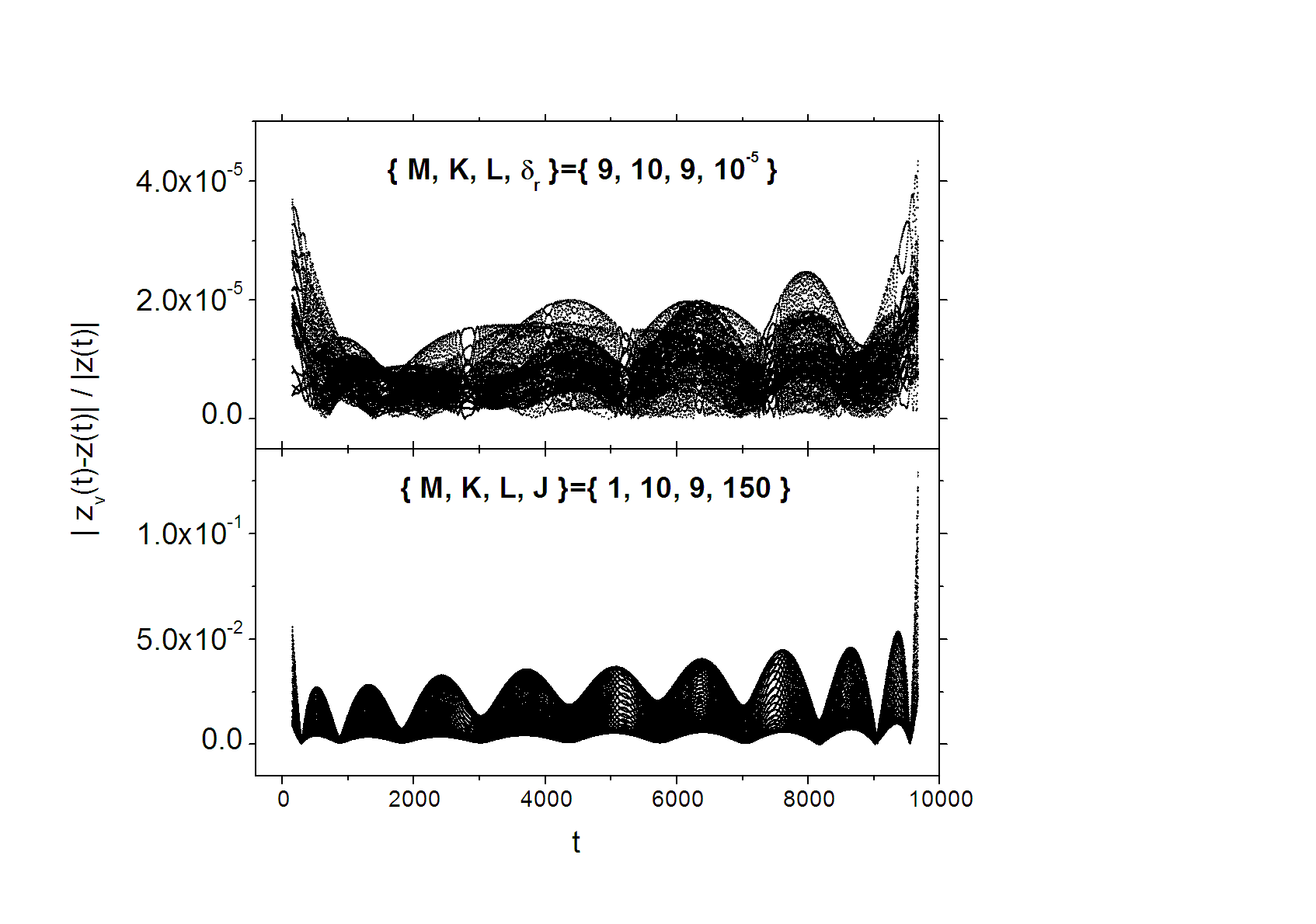}
      \caption{Errors of two different representations of the
dissipated solution specified by (\ref{for D equation}), (\ref{for
D eqnparameter}) and (\ref{for D initialcondition}). In the case of the upper panel, our representation procedure is terminated with a 37-term representation, because this representation reaches the precision requirement $\delta_r=10^{-5}$. In the case of the lower panel, with sligtly larger error in the frequency approximation, the representation is about 4 orders less precise, though the number of representation terms is increased to 150.} \label{fig D precision}
   \end{figure}
}

\newcommand\figureD{
   \begin{figure} 
   \centering
     \includegraphics[width=100mm]{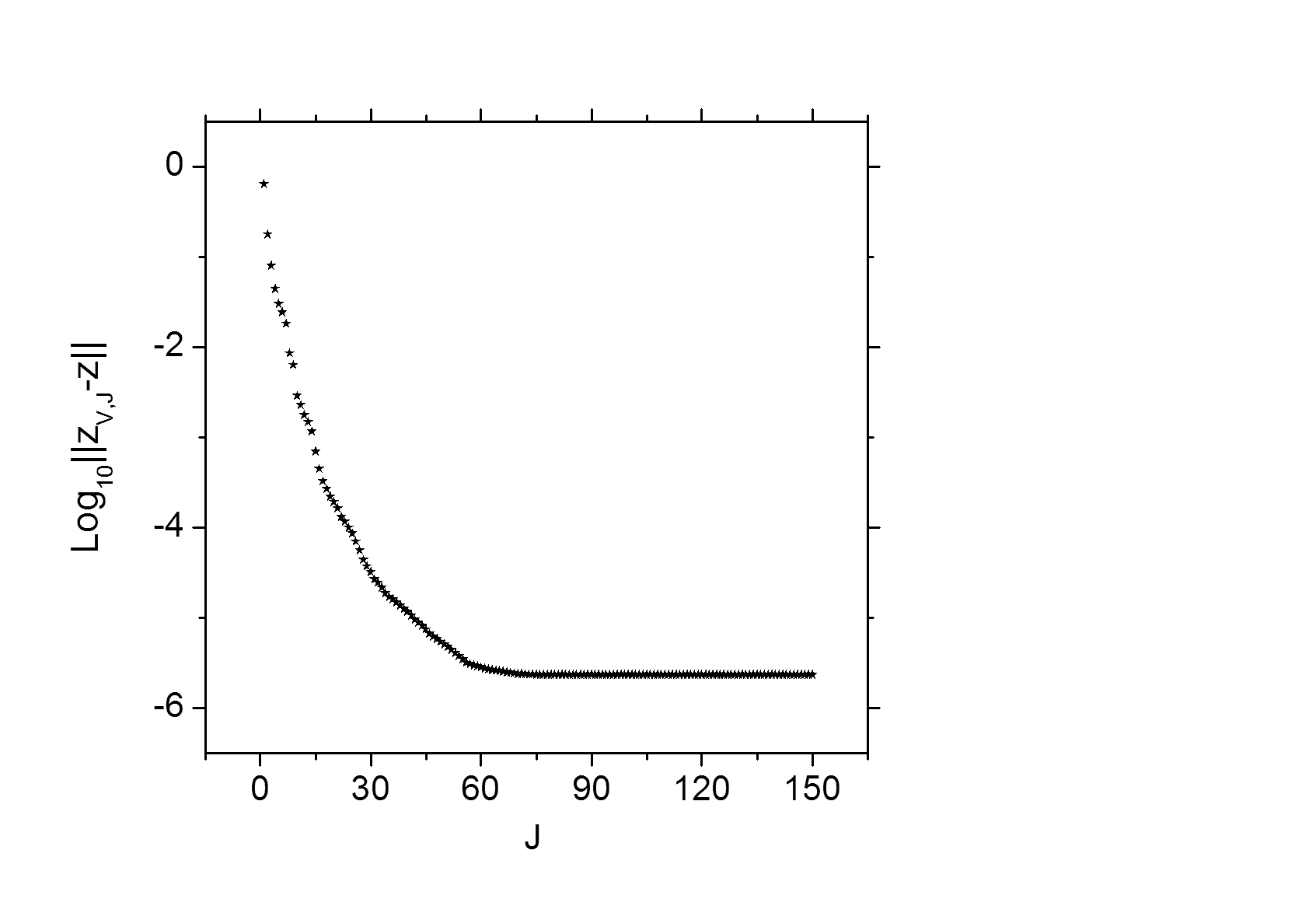}
      \caption{Convergence property of the procedure of representing the dissipated solution specified by (\ref{for D equation}), (\ref{for
D eqnparameter}) and (\ref{for D initialcondition}). The residual of representation is plotted against the number $J$ of representation terms, where other procedure parameters are fixed as $\{M,K,L\}=\{9,10,9\}$. } \label{fig D convergence}
   \end{figure}
}

\newcommand\figureE{
\begin{figure} 
   \centering
   \includegraphics[width=100mm]{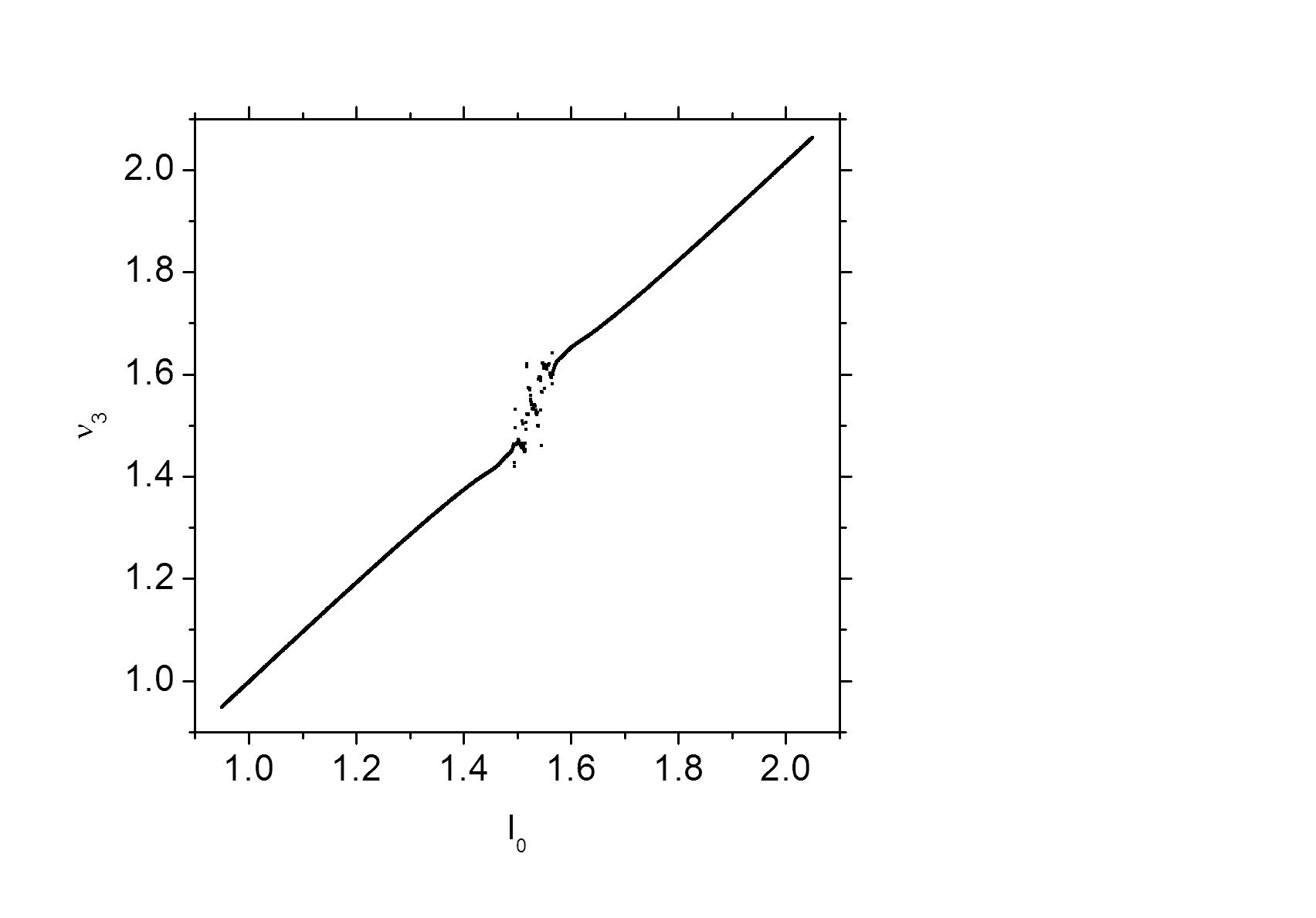}
      \caption{Fundamental frequency map of the system specified by
(\ref{for H equation}) and (\ref{for H eqnparameter}). In the chaotic zone formed by resonance overlap lies the phase point $(\theta_0,I_0)=(0,1.535)$, which will be chosen as the initial phase point of the considered chaotic solution.} \label{fig
CL2 ffmap}
   \end{figure}
}

\newcommand\figureF{
   \begin{figure} 
   \centering
     \includegraphics[width=100mm]{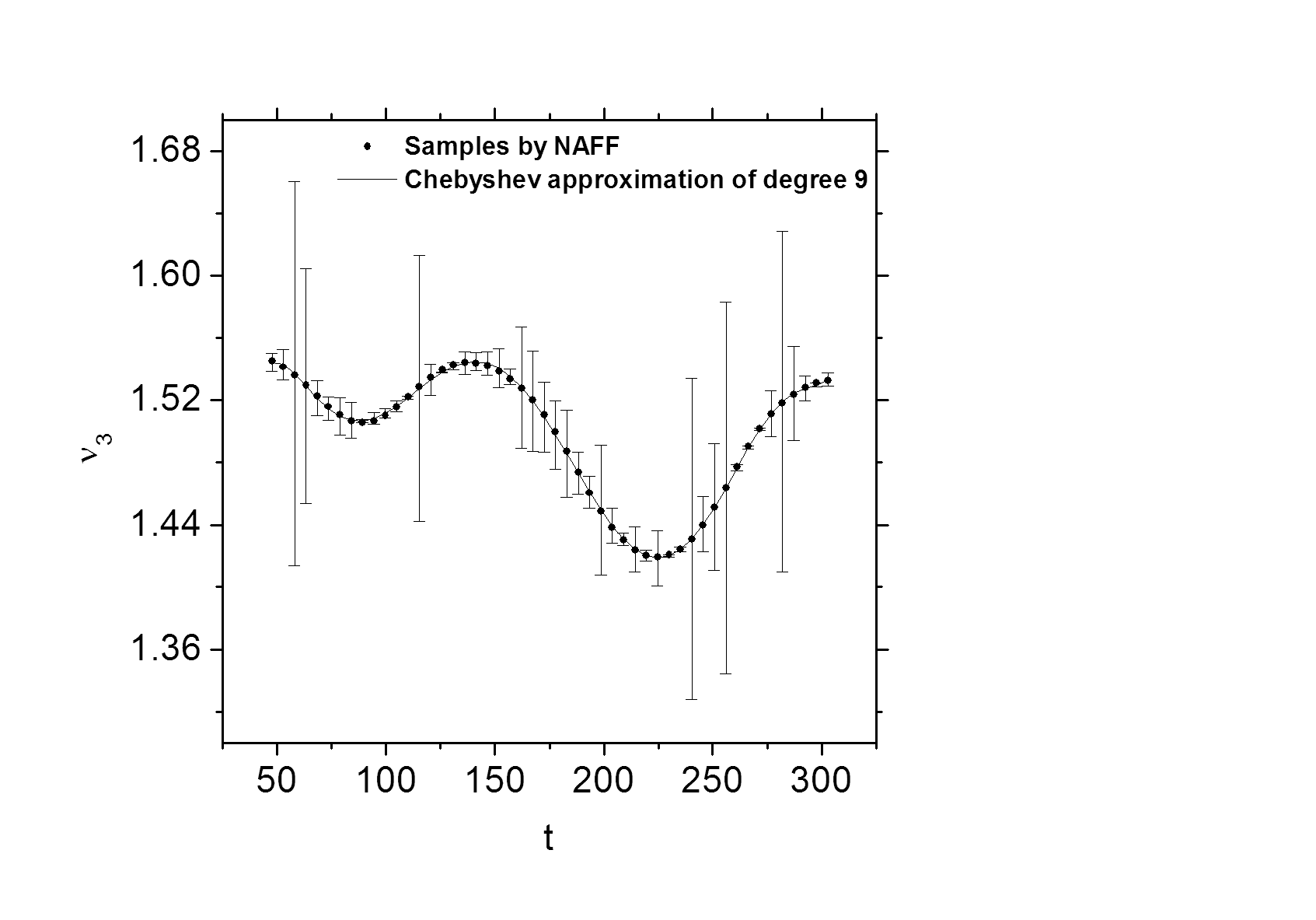}
     \caption{The changing fundamental frequency
$\nu_3(t)$ of the chaotic solution specified by (\ref{for H equation}), (\ref{for H eqnparameter}) and (\ref{for H initialcondition}). The samples of this frequency is computed by applying NAFF algorithm over a sliding time interval, and the errors are estimated as their respective differences from the frequencies of the quasiperiodic approximation of the solution over the same sliding time interval. Also shown is a Chebyshev appoximation of $\nu_3(t)$.} \label{fig CL2 tvsff}
   \end{figure}
}

\newcommand\figureG{
\begin{figure} 
   \centering
     \includegraphics[width=100mm]{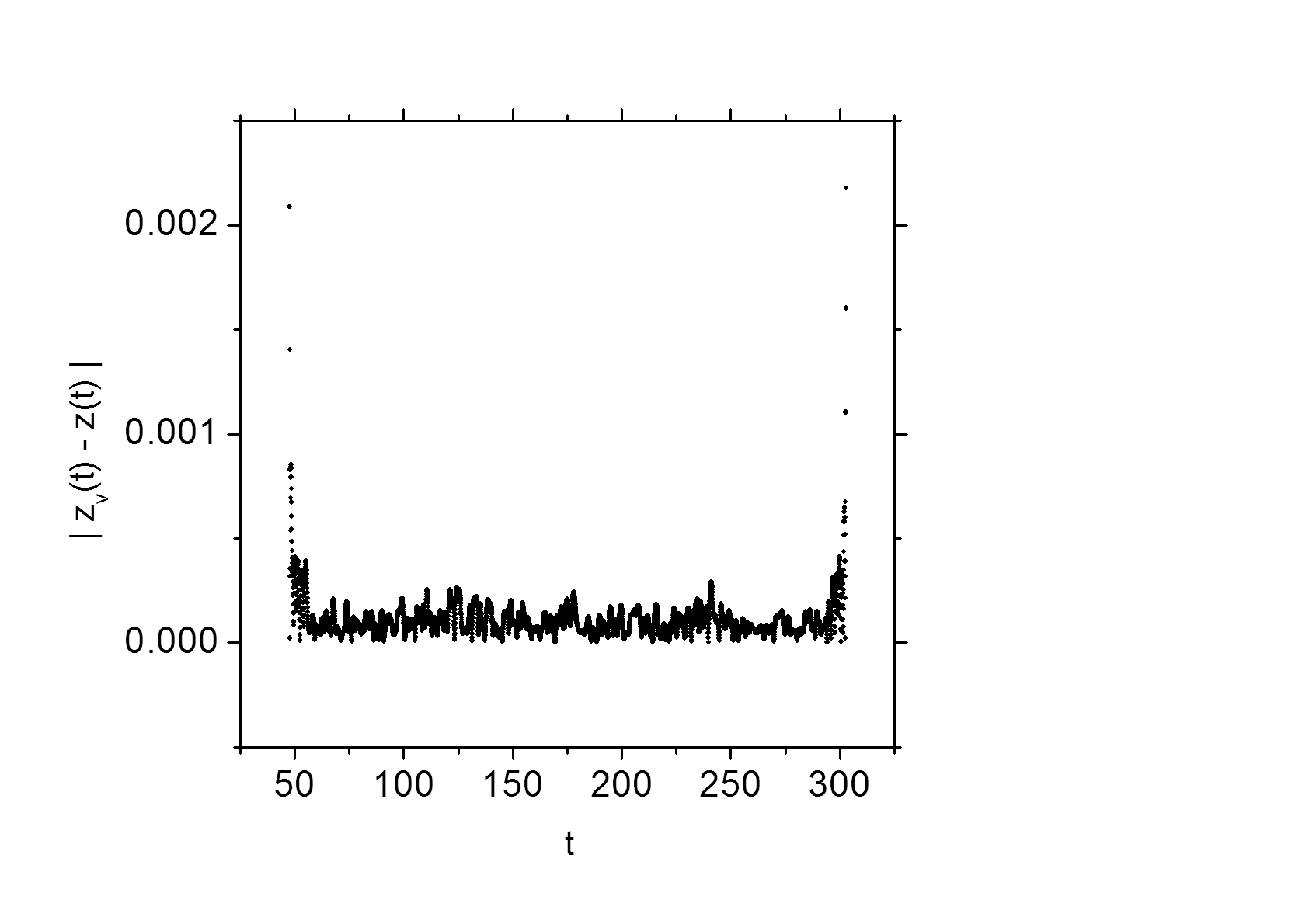}
      \caption{Error of a 100-term representation of the chaotic
solution specified by (\ref{for H equation}), (\ref{for H
eqnparameter}) and (\ref{for H initialcondition}). This representation is obtained with $\{M_1,M_2,M_3,K_1,K_2,K_3,L_\mathbf{k},J\}=\{0,0,9,10,10,10,9,100\}$.} \label{fig CL2 precision}
   \end{figure}
}

\newcommand\figureH{
\begin{figure} 
   \centering
   \includegraphics[width=100mm]{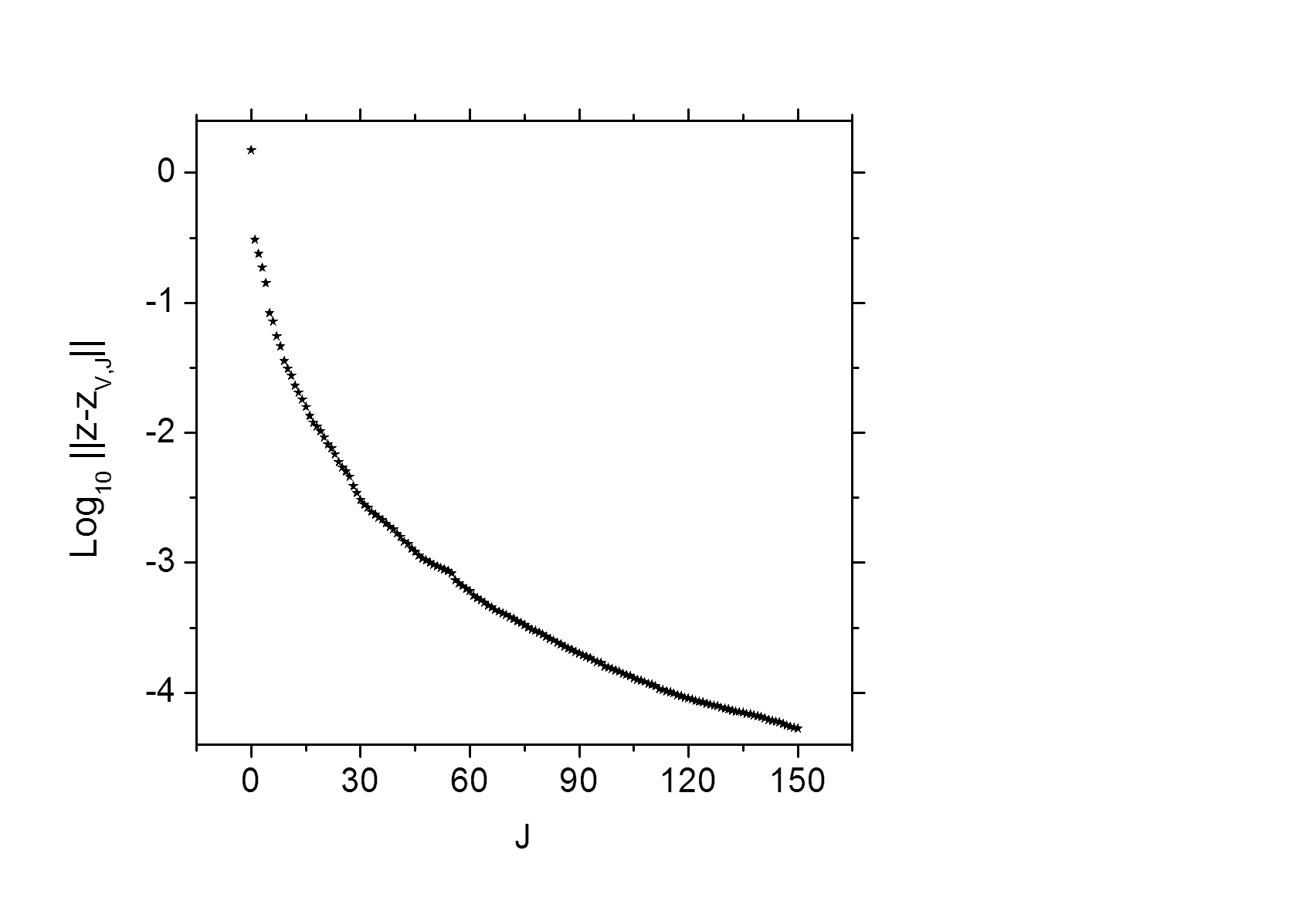}
      \caption{The convergence property of the procedure of representing the chaotic
solution specified by (\ref{for H equation}), (\ref{for H
eqnparameter}) and (\ref{for H initialcondition}). The residual of representation is plotted against the number $J$ of representation terms. The other procedure parameters are fixed as $\{M_1,M_2,M_3,K_1,K_2,K_3,L_\mathbf{k}\}=\{0,0,9,10,10,10,9\}$.} \label{fig CL2
convergence}
   \end{figure}
}

\newcommand\figureI{
\begin{figure*}[h!] 
   \centering
     \includegraphics[width=150mm]{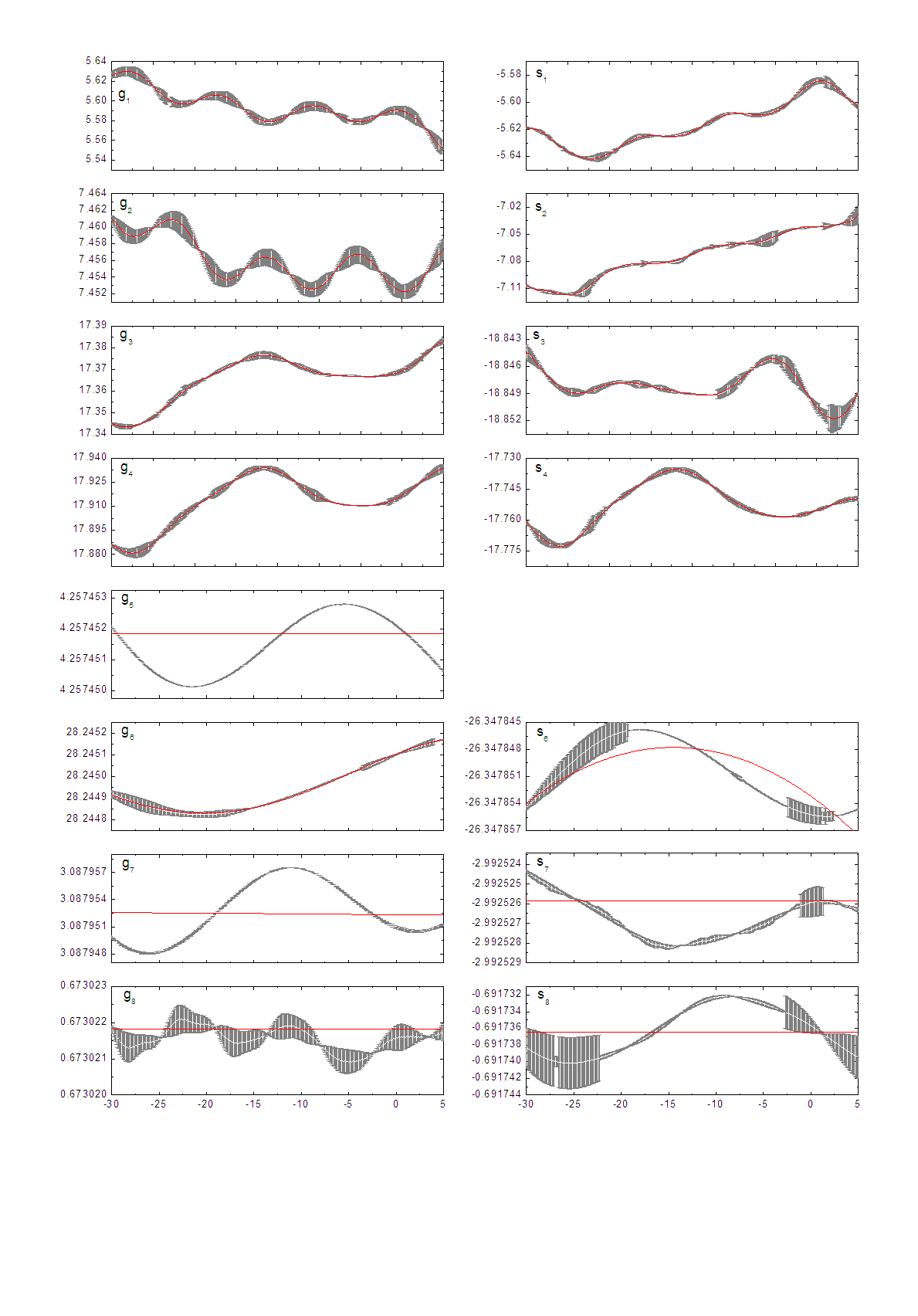}
      \vspace{-3cm} \caption{Variations of major fundamental frequencies of the solar system  over the time interval from -35 Myr to 5 Myr with origin at J2000. These frequencies (in arcsec\ yr$^{-1}$) are computed by applying the NAFF algorithm to the proper modes of the secular solar system associated respectively with major planets. The errors in the resulting frequency samples are estimated as the difference between the frequencies computed respectively from the original ephemeris and its quasiperiodic approximation. Also shown in this figure are the Chebyshev approximations of these varing fundamental frequencies (red curves), which are specified in Table \ref{tab FFChC}.} \label{fig solar FFs}
\end{figure*}
}

\newcommand\figureJ{
\begin{figure}[h!] 
   \centering
     \includegraphics[width=85mm]{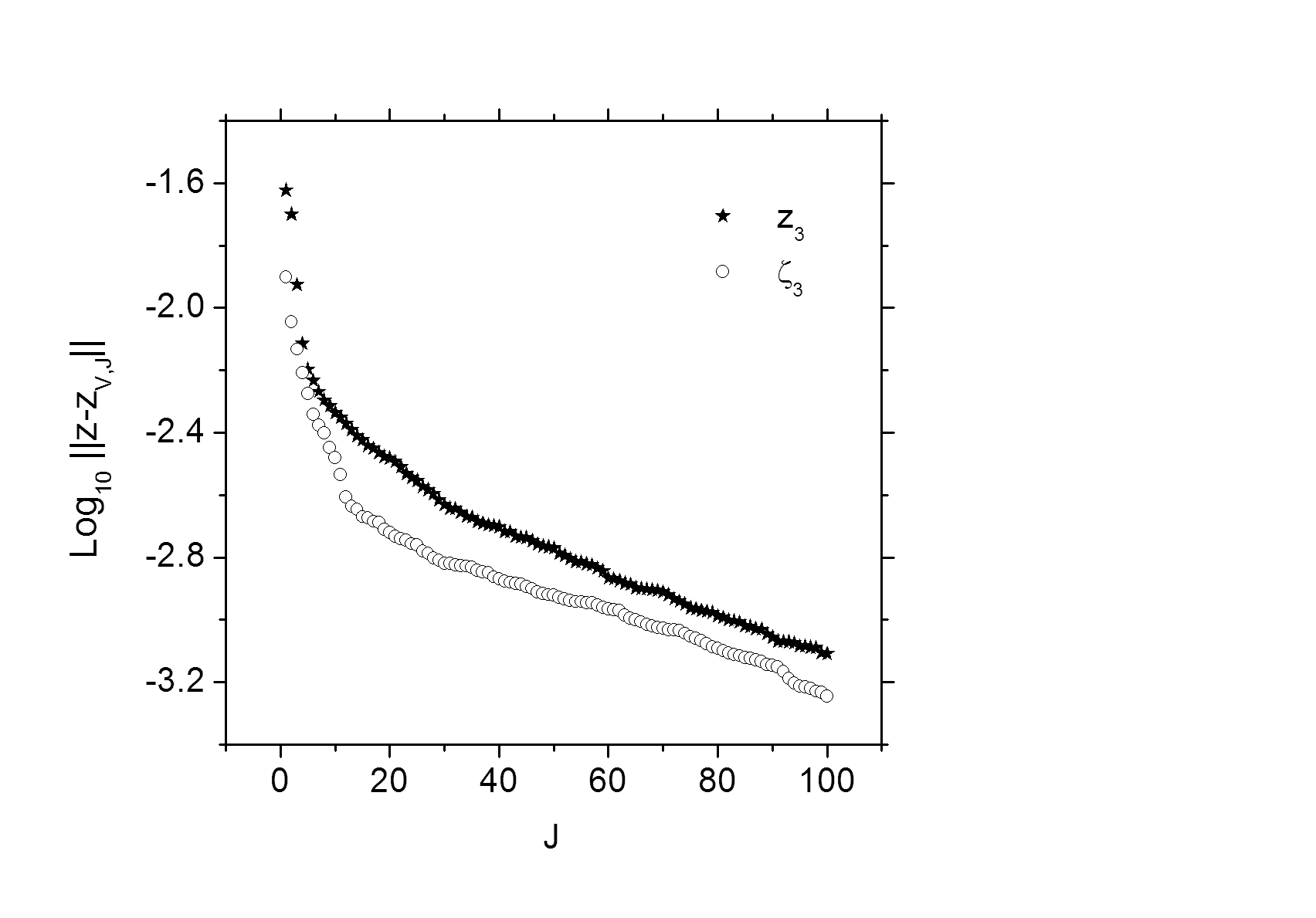}
      \caption{Convergence property of the two representation procedures leading respectively to the 100-term representations of $z_3(t)$ and $\zeta_3(t)$. The residual of representation is plotted against the number $J$ of representation terms.} \label{fig convergencela04}
\end{figure}
}

\newcommand\figureK{
\begin{figure*}[h!]  
   \centering
     \includegraphics[width=150mm]{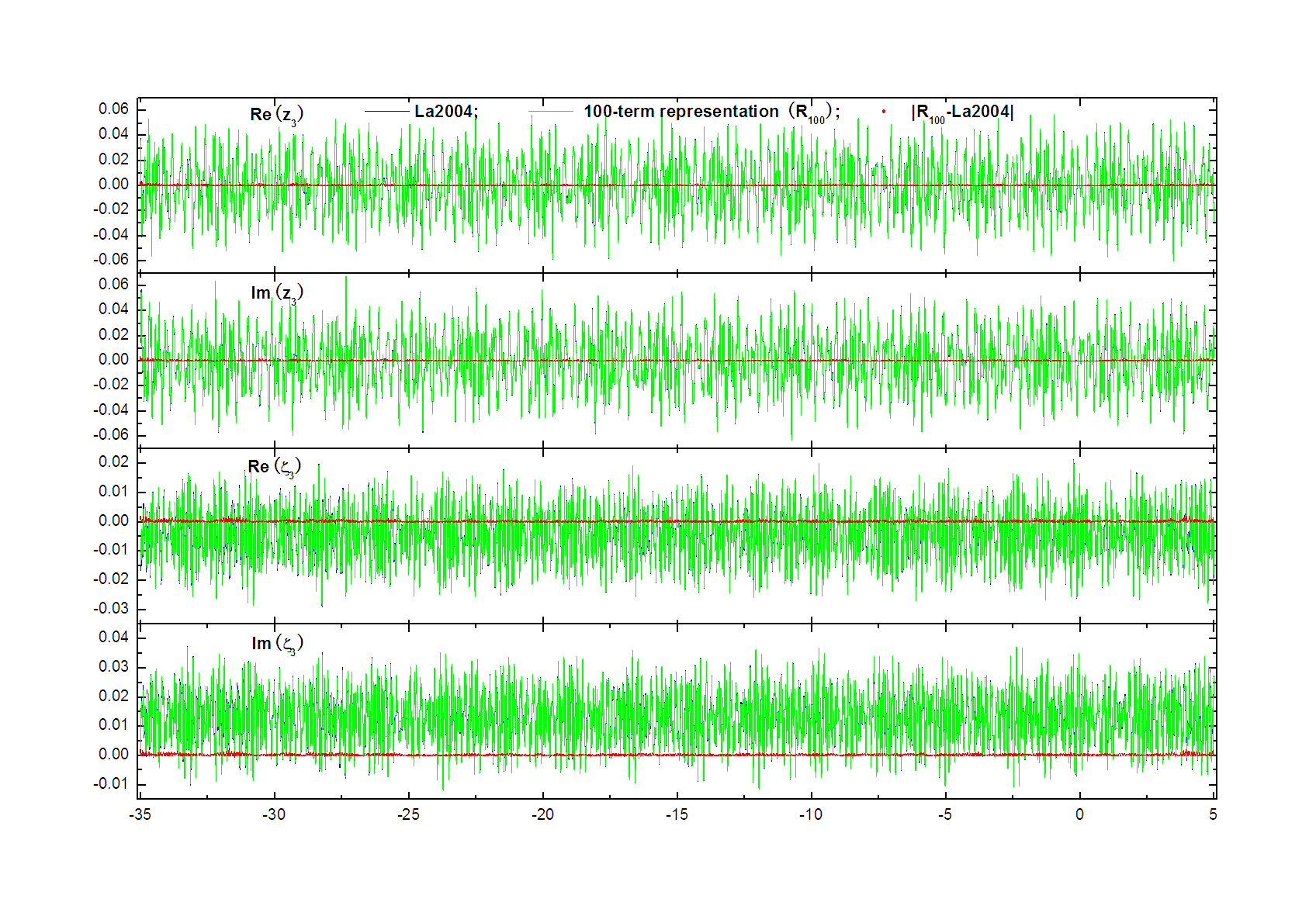}
      \caption{Comparison between the eccentricity and inclination ephemerides of La2004 and their respective 100-term representations over the time interval from -35 Myr to 5 Myr with origin at J2000. Both representions are in the form of (\ref{z zata 100}), with the Chebyshev expansions approximating the fundamental frequencies specified in Table \ref{tab FFChC}, and main frequency index vectors and complex amplitudes of all 100 terms in the on-line electronic files under names z3R100.dat and zeta3R100.dat, respectively.} \label{fig EphvsRepla04}
   \end{figure*}
}

\newcommand\figureL{
\begin{figure*}[h!]  
   \centering
     \includegraphics[width=150mm]{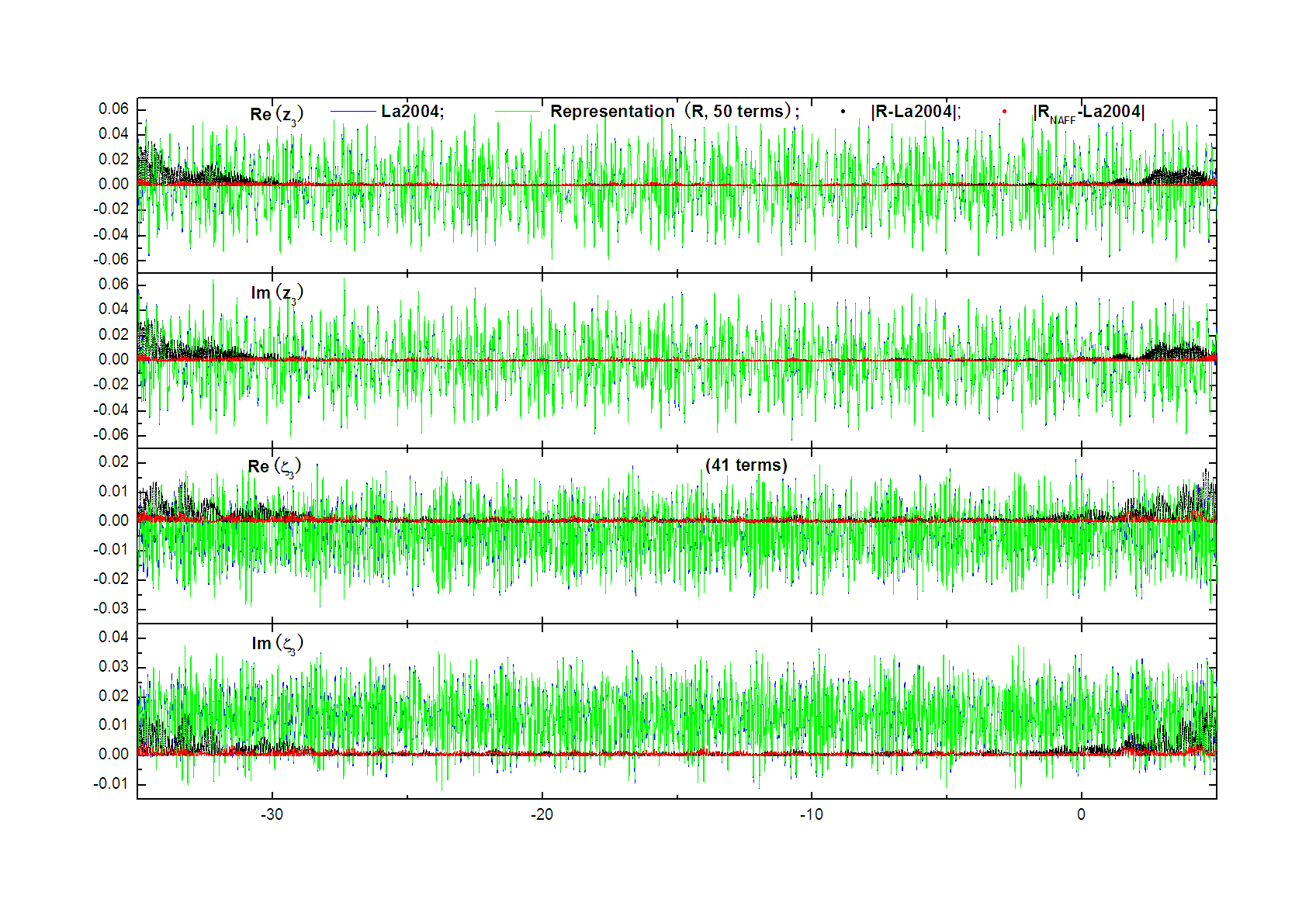}
      \caption{Comparison between the eccentricity ephemeris $z_3$ and the inclination ephemeris  $\zeta_3$ of La2004 and their respective representations by NAFF ($\Re_{NAFF}$) and by the present algorithms ($\Re$). The number of terms of $z_3$-representation is 50 for both $\Re_{NAFF}$ and $\Re$, and the number is 41 in the case of  $\zeta_3$. \label{fig VqpQpla04}}
   \end{figure*}
}

\begin{document}
   \title{Frequency analysis and the representation of slowly diffusing  planetary solutions}


   \author{Y.N. Fu
          \inst{1}
          \and
          J. Laskar\inst{2}
          }

   \offprints{J. Laskar}

   \institute{Purple Mountain Observatory, Chinese Academy of Sciences, 2 West Beijing Road, Nanjing 210008, China\\
              \email{fyn@pmo.ac.cn}
         \and
             ASD, IMCCE-CNRS UMR8028, Observatoire de Paris, UPMC, 77 Av. Denfert-Rochereau, 75014 Paris, France\\
             \email{laskar@imcce.fr}
             }
   \date{Received ; accepted}


  \abstract
   {Over short time intervals planetary ephemerides have been traditionally represented in analytical form
   as finite sums of periodic terms or sums  of Poisson terms that are periodic terms with polynomial amplitudes.
   Nevertheless, this representation is not well adapted for the  evolution of the planetary orbits in the solar system
   over million of years  as they  present  drifts in their main frequencies, due to the chaotic nature of their dynamics. }
   {The aim of the present paper is to develop a numerical algorithm for   slowly diffusing
   solutions of a perturbed integrable Hamiltonian system that will apply to the representation of the
   chaotic planetary motions with varying frequencies.}
   {By simple analytical considerations, we first argue that it is possible to recover exactly
   a single varying frequency. Then, a function basis involving time-dependent fundamental frequencies
   is formulated in a semi-analytical way. Finally, starting from a numerical solution, a recursive algorithm
   is used to numerically decompose the solution on the significant elements of the function basis.}
   {Simple examples show that this algorithm can be used to give compact representations of different
   types of slowly diffusing solutions. As a test example, we show how this algorithm can be successfully applied to
   obtain a very compact approximation of the La2004 solution of the orbital motion of the Earth over 40 Myr ([-35Myr,5Myr]).
   This example has been  chosen as this solution is widely used for the reconstruction of the climates of the past.}
   {}


\authorrunning{Fu \& Laskar}
\titlerunning{Frequency analysis \& solution representation}

   \maketitle
%

\section{Introduction}
Before the computer ages, the long term
solutions for the planetary orbits were derived by perturbation methods,
and obtained in the form of sum  of periodic terms. The first of such solutions was obtained by
\citet{Lagr1782a} who only considered  the known planets of the time, that is the planets that are visible by naked eye
(Mercury, Venus, Earth, Mars, Jupiter and Saturn).
This was later on improved by \citet{LeVe1840a,LeVe1841a} who took
also Uranus  into account. These long term solutions have revealed to be of
fundamental importance for the understanding of the past climate of the Earth,
when it was understood that the changes of the orbit of the Earth induce  also some change in its obliquity
and  in the insolation at the Earth surface \citep{Mila1941a} (for a detailed review, see
\citet{laskar04}).
With the advent of computers, two different approaches become possible.
Computer algebra allowed to extend the perturbation methods \citep[e.g.][]{Bret1974a}, and
as the computer speed increased, direct integrations of the full Solar system become
possible \citep{QuinTrem1991a,SussWisd1992a}. Meanwhile, \citet{Lask1985a,Lask1986a,laskar88} developed a mixed strategy,
with an analytical averaging of the planetary equations obtained by perturbation methods  using dedicated computer algebra.
This analytical averaging was then   followed by
a numerical integration of the averaged system with high order multistep method.
In order to compare the output of the numerical integrations
with the quasiperiodic  solutions of the perturbative methods, \citet{laskar88} introduced the frequency analysis method
that allowed  to obtain in a very efficient way a precise approximation of  the numerical solution in
quasiperiodic form (e.g.  \citet{laskar03}).

But although the goal was to search for the most precise
long term solution for the planetary orbits,
one of the outcomes of these computations was  to demonstrate that the solar system motion
is chaotic \citep{Lask1989a,laskar90}. As a consequence, the solutions are not quasiperiodic,
although they can be approximated by a quasiperiodic expression over a limited time of a few million of years
\citep{laskar88,laskar90,laskar04}.

In the present work, we derive a more adapted strategy for the slowly diffusing trajectories of a dynamical system
that will be  very well suited to the  construction of compact forms for the long time
behaviour of the planetary orbits.
We introduce an algorithm, that is derived from the frequency  analysis,
but where a slow variation of the frequencies and amplitudes is  added.
Here, by a slowly diffusing solution, we mean a solution that, while
experiencing significant frequency drifts in the whole considered time interval, is
nearly quasiperiodic in time subintervals. Similar to frequency analysis, a key step of our algorithm is to
construct frequency-dependent function basis on which the considered solution
will be decomposed.

In section 2, after reminding some fundamental results on frequency analysis,
we consider a single term model and show that it is
possible to recover exactly its varying frequency as a function of time. This
is important when one is interested in the details of how a solution diffuses
in the frequency space. But for constructing a compact solution representation
with a reasonably high precision, small flickers
without significant cumulative effects can be smoothed out from a varying
frequency. Therefore, it is preferable to use a  model with
limited number of parameters, e.g. low-order polynomials, to approximate the
frequency. To do so, we  sample   the average frequencies over a
sliding time interval. For a slowly diffusing solution, it
is assumed that all fundamental frequencies can be sampled in this way by using
the NAFF algorithm (e.g. \citep{laskar03}).

In section 3, a general algorithm of representing a slowly diffusing solution
is designed, where a frequency-dependent function basis is constructed based on
the Chebyshev approximations of fundamental frequencies. The basis functions
with significant but non-fundamental frequencies are generated according to the
assumption that, at any instant, a slowly diffusing solution is
nearly quasi-periodic with the smoothed fundamental frequencies at that
instant, namely all main frequencies are integral linear combinations of the
fundamental frequencies. This effectively avoids the difficulties in determining
the frequencies of long-period terms and/or groups of neighboring-period terms.

Two simple examples are given in section 4. In the first example, a
weakly dissipated system is considered, and the represented solution diffuses
because of dissipation. In the second example, a Hamiltonian system of
degree 1.5 is considered, and a solution starting from an obvious resonance
overlap zone is represented. To show the flexibility of our algorithm in
representing such a solution, we exclude all possible libration frequencies
from the set of fundamental frequencies used in constructing the function basis.

Applications to the representation of ephemerides of the solar system
bodies are provided in section 5. As examples, the eccentricity and the inclination
of the Earth, as given by the long-term numerical solution La2004
\citep{laskar04}, are represented. For such a
realistic solution, it is natural to take into account the restrictions
on the representation model from previous results, e.g. the important
libration frequencies from numerical analysis \citep[e.g.][]{laskar04}, so that the
resulting representations can be as close  as possible to the physical model.

\section{Frequency analysis and time-dependent frequencies}

For a KAM solution of a dynamical system  \citep{kolmogorov,arnold,moser}, with fundamental
frequency vector $\mathbf{\nu}=(\nu_1,...,\nu_N)\in R^N$,
where $N$ is the number of degrees of freedom of the system, the motion
of any given degree of freedom, with associated  variable $z(t)$, can be described in
complex form as

\begin{equation}
\label{for qp model} z(t)=\sum_{\mathbf{k} \in Z^N} a_{\mathbf{k}}
\e^{\i \langle \mathbf{k},\mathbf{\nu} \rangle t},
\end{equation}
where $\mathbf{k}=(k_1,...,k_N)$ is the frequency index vector,
$a_\mathbf{k}$ complex amplitude, and $\langle \cdot, \cdot
\rangle$ denotes the usual inner product of two real vectors.

The NAFF (Numerical Analysis of Fundamental Frequencies) algorithm was designed to numerically recover significant
terms of (\ref{for qp model}) from a numerical sample set of $z(t)$,
e.g. an ephemeris obtained by numerical integration
\citep{laskar88}. However, the application
of NAFF is not limited to numerically recovering KAM solutions.
Actually, most relevant works are about frequency drifts. In
particular, the variations of the fundamental frequencies of the
secular solar system over 200Myr was used to study the chaotic
behavior of the solar system   \citep{laskar90}.

Let us first recall briefly the theorem about the convergence of the
NAFF algorithm (for the complete version with proof, see
\citep{laskar03}). With a set of $N$ appropriate variables, the complex
function (\ref{for qp model}) describing a degree of freedom of an analytic
KAM solution can be written as

\begin{equation}
\label{for KAM solution} z(t)=\e^{\i
\nu_1t}+\sum_{\mathbf{k} \in Z^N-(1,0,...,0)}a_\mathbf{k}
\e^{\i \langle \mathbf{k},\mathbf{\nu} \rangle t}\ \ \ \
(|a_{\mathbf{k}}|<1),
\end{equation}
where $\nu_1$ is a fundamental frequency. We have then
\citep[e.g.][]{laskar99}

{\bf Theorem 1} Let $\nu_1^T$ be the value of $\sigma \in
R$ that maximizes the function $\phi(\sigma)=|\langle
z(t),\e^{\i \sigma t} \rangle_T^{ \chi} |,$ where $\chi =\chi
(t/T)>0$ is a weight function, and $\langle \cdot,\cdot
\rangle_T^\chi$ the inner product of two complex functions of $t$
defined as
\begin{equation}
\label{for innerproduct} \langle f(t),g(t) \rangle_T^{ \chi } =
\frac{1}{2T}\int_{-T}^T f(t)\bar{g}(t) \chi(t/T)dt,
\end{equation}
we then have $\lim_{T\rightarrow \infty}\nu_1^T=\nu_1$.$\square$

Similar to the fundamental frequency $\nu_1$, any other main
frequency can be recovered by searching its neighborhood or
$R$ for the value of $\sigma$ that maximizes $\phi(\sigma)$ defined
with the remaining $z(t)$, that is, the original $z(t)$ minus the
recovered terms.

A diffusion solution is then characterized by fundamental frequency variations
\citep{laskar92,laskar93}.
Regarding to the determination of a varying frequency, let's first consider the following simplest case,
\begin{equation}
\label{for one term qp func} f(t)=a(t)\e^{\i \int_0^t
\nu(\tau)d \tau},
\end{equation}
where $\nu(t)$ and $a(t)>0$ are real integrable functions. Writing
\begin{equation}
\phi(\sigma(t))=\left|\langle
f(t),\e^{\i\int_0^t\sigma(\tau)d\tau}\rangle_T^\chi\right|
=\left|\frac{1}{2T}\int_{-T}^Ta(t)\e^{\i\theta(t)}\chi(t/T)dt\right|,
\label{for v power1}
\end{equation}
where $\theta(t)=\int_0^t(\nu(\tau)-\sigma(\tau))d\tau$ belongs to $\mathcal{C}^0$, the set of continuous function on $R$,
we have the following theorem

{\bf Theorem 2} For any given $T>0$ and $\nu(t) \in \mathcal{C}^0$, the functional $\phi(\sigma(t))$ has one and only one maximum in $\mathcal{C}^0$, which is attained at
$\sigma(t)\equiv\nu(t)$. $\square$

Proof. The right-hand side of (\ref{for v power1}) can be written as
$$
\left|\frac{1}{2T}\int_{-T}^{T} (\sqrt{a(t)}) (\sqrt{a(t)}\e^{\i\theta(t)})\chi(t/T) dt\right|=
\left|\langle \sqrt{a(t)},\sqrt{a(t)}\e^{-\i\theta(t)} \rangle_T^{ \chi}\right|.
\label{Preposition1}
$$
By using the Cauchy-Bunyakovsky-Schwarz inequality\footnote{Given any two vectors $f$ and $g$ in an inner product space, the Cauchy-Bunyakovsky-Schwarz inequality writes $\langle f,g \rangle\leq \|f\| \times \|g\|$, where $\langle \cdot,\cdot \rangle $ denotes inner product and $\|\cdot\|$ the induced norm. Also, the equality is true if and only if $f$ and $g$ are linearly dependent.}, it is easy to deduce from this formula that
$$
\phi(\sigma(t))\leq \|\sqrt{a(t)}\| \times \|\sqrt{a(t)}\e^{-\i\theta(t)})\|=\frac{1}{2T}\int_{-T}^{T}a(t)\chi(t/T)dt
$$
where $\|\cdot\|$ denotes the norm induced by the inner product (\ref{for innerproduct}). Moreover, the equality is true if and only if $\sqrt{a(t)}$ and $\sqrt{a(t)}\e^{-\i\theta(t)}$ are linearly dependent, i.e. $\theta(t)$ is a constant. This constant is 0, since $\theta(0)=0$ by definition. The above arguments imply that $\phi(\sigma(t))$ has a unique maximum attained at $\theta(t)\equiv 0$. Since $\nu(t)$ is continuous and the searched $\sigma(t)$ is also continuous, $\theta(t)\equiv 0$ is equivalent to $\sigma(t)\equiv\nu(t)$. Therefore, it is concluded for any given $T>0$ that
$\phi(\sigma(t))$ has one and only one maximum attained at
$\sigma(t)\equiv\nu(t)$.$\square$

Theorem 2 implies that, even in the case of a varying frequency, it is still possible to recover exactly the frequency, as a function of time.
Now, let us consider the following more general complex function
\begin{equation} \label{for qp func} f(t)=a_1(t) \e^{\i
\int_0^t\nu_1(\tau) d\tau}+\sum_{\mathbf{k} \in
Z^N-(1,0,...,0)}a_\mathbf{k}(t) \e^{\i (\theta_{\mathbf{k}} +
\int_0^t \langle \mathbf{k},\mathbf{\nu}(\tau) \rangle d \tau)},
\end{equation}
where $a_1 (a_{\mathbf{k}}) \in S_a$ with $S_a$ a subspace of real
function space, $\theta_{\mathbf{k}} \in
R$, and $\{\nu_n\}_{n=1}^N \in S_{\nu}$ with $S_{\nu}$ a linear
subspace of the real integrable function space. As an extension of
(\ref{for KAM solution}), this function inherits an
important time varying character of (\ref{for KAM solution}), i.e. all phase
increments are described by a single fundamental frequency vector
$\mathbf{\nu}(t)$. We will make the heuristic assumption that, under suitable conditions, a similar result as Theorem 1 holds for more general expression with varying frequencies as in (\ref{for qp func}).

\textit{Assumption : If $\|\nu_1^T-\nu_1\|$ is sufficiently small, where
$\nu_1^T \in S_{\nu}$ and $\|\cdot\|$ is an appropriately defined
$T$-dependent norm (e.g. the one induced by the inner product
(\ref{for innerproduct})), and $\sigma(t)=\nu_1^T(t)$ (locally)
maximizes the functional $\phi(\sigma)=|\langle f(t),\e^{\i
\int_0^t \sigma(\tau) d\tau } \rangle_T^{ \chi} |$, then
$\lim_{T\rightarrow \infty}\nu_1^T(t)=\nu_1(t)$.}

\section{Representation of slowly diffusing solutions}

From now on, we will restrict ourselves to slowly diffusing
solutions of an ordinary differential equation system obtained by
slightly perturbing an integrable Hamiltonian system. We denote by
$(I,\theta)=\{I_n,\theta_n\}_{n=1}^N$ the action-angle variables
of the integrable Hamiltonian system, which will be used to
express the ephemeris of a diffusing solution
$\{z_n(t)\equiv I_n(t) \e^{\i \theta_n(t)}\}_{n=1}^{N}$.

\subsection{Representation procedure}

We start from a sample set of $\{z_n(t)\}_{n=1}^N$. Suppose that
the samples are given, respectively, at grid points from $t_0$ to $t_1$
with fixed time step $h$, which is much smaller than the minimum of the
fundamental periods. In accordance with the condition of slow
diffusion, we assume that the solution is close to quasi-periodic
in below mentioned time subintervals (of $[t_0,t_1]$) with length much larger than the
maximum of the fundamental periods. These roughly stated preconditions are required because we will
use NAFF algorithm to estimate the changing fundamental frequencies.

As the first step, we apply NAFF algorithm to obtain fundamental frequency samples. For each given degree ($n$),
this algorithm will be applied, respectively, to $z_n(t)$ samples over evenly spaced time
subintervals $\{[\tau_\lambda-\frac{d}{2},\tau_\lambda+\frac{d}{2}]\}_{\lambda=1}^{\Lambda}$
of $[t_0,t_1]=[\tau_1-\frac{d}{2},\tau_\Lambda+\frac{d}{2}]$. For each of these subintervals, the first recovered frequency will be taken as the averaged value of the fundamental
frequency $\nu_n$ over the same subinterval. The $N$ averaged fundamental frequencies obtained in this way are
then taken as the instant ones at $\tau_\lambda$, resulting
in the fundamental frequency samples $\{\tau_\lambda,\nu_n(\tau_\lambda)\}_{\lambda=1}^\Lambda, n=1,...N$.

Secondly, we fit for each given degree the frequency samples to a Chebyshev
expansion valid on $[\tau_1,\tau_\Lambda]$,
\begin{equation}
\label{for nu Cheb} \nu_n(t)=\sum_{m=0}^{M_n}c_{m,n}
T_m(x)\ \ \ \ \ \ \ \ \ \ \ \ \ (n=1,...,N),
\end{equation}
where $c_{m,n} \in
R$ is a Chebyshev coefficient, $x=\frac{2(t-\tau_1)}{(\tau_{\Lambda}-\tau_1)}-1 \in [-1,1]$
a normalized time, and $T_m(x)$ the Chebyshev polynomial\footnote{The explicit expressions of
$T_m(x)$ up to $m=15$ are given  in (\ref{for 15 Chebyshev polynomials}).}
 of degree $m$.  We then construct numerically a frequency-dependent function
basis $B$ (see the next subsection), on which the considered solution will be decomposed.

The final step, i.e. decomposing $z_n(t)$ on $B$, is the
same as that of the NAFF algorithm \citep{laskar99}, except for
the searched function bases for significant terms. The function
bases are $\{ \e^{\i \omega_k t}, \omega_k \in
R \}$ for NAFF and $B$ for the presently described procedure.

\tableA

\subsection{Representation model}
By variation of parameters, (\ref{for qp model}) becomes
\begin{equation}
\label{for v model 1} z_{v}(t)=\sum_{\mathbf{k} \in
Z^N} a_{\mathbf{k}}(t) \e^{\i \int_0^t\langle
\mathbf{k},\mathbf{\nu}(\tau) \rangle d\tau}.
\end{equation}
Now, let
\begin{equation}
\label{for frequency index set}
\hat{Z}^N = \{k_n: |k_n|\leq K_n \in Z^+ \}_{n=1}^N
\end{equation}
be a truncated set of the frequency index vector $\mathbf{k}$. And, for each $\mathbf{k} \in \hat{Z}^N$, let
\begin{equation}
\label{for Cheb app of amplitude}
a_{\mathbf{k}}(t)\approx\sum_{l=0}^{L_\mathbf{k}} \widetilde{a_{l,\mathbf{k}}}\ T_l(x(t))
\end{equation}
\noindent be a Chebyshev expansion of the amplitude valid on $[\tau_1,\tau_{\Lambda}]$, where
$\widetilde{a_{l,\mathbf{k}}} \in
C$.
These lead us from (\ref{for v model 1}) to the following representation model of $z_v(t)$
\begin{equation}
\label{for v model 2} z_v(t)\approx\sum_{\mathbf{k} \in \hat{Z}^N}
\sum_{l=0}^{L_\mathbf{k}} \widetilde{a_{l,\mathbf{k}}}\ T_l(x(t)) \e^{\i
\int_0^t\langle \mathbf{k},\mathbf{\nu}(\tau) \rangle d\tau}.
\end{equation}

From the Chebyshev approximations of the fundamental frequencies, it is
easy to obtain by integration the Chebyshev expansion representing
the phase increment associated with the frequency $\langle
\mathbf{k},\mathbf{\nu}(\tau) \rangle$, from the value, as is preferred, at the middle of the time interval $[\tau_1,\tau_\Lambda]$,
\begin{equation}
\label{for Phi}
\int_{(\tau_1+\tau_\Lambda)/2}^t\langle
\mathbf{k},\mathbf{\nu}(\tau) \rangle d\tau=\varphi_{0\mathbf{k}}+\varphi_\mathbf{k}(t) \ \ \ (\mathbf{k} \in \hat{Z}^N)
\end{equation}
where $\varphi_{0\mathbf{k}} \in
R$, and $\varphi_\mathbf{k}(t)$ gathers all Chebyshev polynomials of
degree larger than 0. With $\varphi_{0\mathbf{k}}$ and
$\varphi_\mathbf{k}(t)$, (\ref{for v model 2}) can be written as
\begin{equation}
\label{for v model 3} z_v(t)\approx \sum_{\mathbf{k} \in
\hat{Z}^N} \sum_{l=0}^{L_\mathbf{k}} a_{l,\mathbf{k}}\ T_l(x(t)) \e^{\i
\varphi_\mathbf{k}(t)},
\end{equation}
where $a_{l,\mathbf{k}}=\widetilde{a_{l,\mathbf{k}}}
\e^{\i\varphi_{0\mathbf{k}}} \in C$ is simply the coordinate of
$z_v(t)$ when it is decomposed on the function basis
\begin{equation}
\label{for v basis} B=\{b_{l,\mathbf{k}}\}=\{T_l(x(t)) \e^{\i
\varphi_\mathbf{k}(t)}\}.
\end{equation}

For the obtained representation, it should be noted that, though we
decompose a solution on the basis $B$, and correspondingly, express
its representation by (\ref{for v model 3}), $\varphi_\mathbf{k}(t)$
will not be used to specify the representation,
$\{\nu_n(t)\}_{n=1}^N$ will be used instead. In other words, the
representation will be specified by the coordinates
$a_{l,\mathbf{k}}$, and the Chebyshev coefficients of
$\{\nu_n(t)\}_{n=1}^N$. This choice is made for the following two
reasons. One is that the representation could otherwise be unnecessarily
cumbersome. The other reason is that, from the Chebyshev coefficients of
$\{\nu_n(t)\}_{n=1}^N$, those of $\varphi_\mathbf{k}(t)$ can be
easily obtained according to (\ref{for Phi}). Here, one should
also be reminded of the fact that the solution representation,
because of its dependence on the Chebyshev approximations of $\nu_n(t)$
and $a_\mathbf{k}(t)$, is valid on $[\tau_1,\tau_{\Lambda}]$ rather than
$[t_0,t_1]$.

To complete the description of the representation model, we point out
that the integers $\{M_n,K_n\}_{n=1}^N$ and $\{L_\mathbf{k}\}_{\mathbf{k}\in\hat{Z}^N}$,
introduced respectively in (\ref{for nu Cheb}), (\ref{for frequency index set}) and (\ref{for Cheb app of amplitude}),
are necessary parameters for defining a particular representation procedure.
Of course, the number of these parameters can be reduced by requiring that
some or all of these parameters take the same sufficiently large value, e.g.
$L_\mathbf{k}=L$ for all $\mathbf{k} \in \hat{Z}^N$, at the price of unnecessarily increasing the basis dimension. Another important point to mention is that,
for terminating the procedure, one needs to set beforehand the maximum number of representation terms ($J$)
and/or the required precision. The required precision are specified
by using either the absolute truncation error $\delta$ or the relative
truncation error $\delta_r$. Correspondingly, the procedure will be
terminated if

\begin{equation}
\label{for absolute error}
\|z-z_v\|<\delta
\end{equation}
or
\begin{equation}
\label{for relative error}
\|z-z_v\|/\|z\|<\delta_r,
\end{equation}
where the module $\|\cdot\|$ is induced by the inner product (\ref{for
innerproduct}) with prescribed $\chi \equiv 1$. In the following, the above-mentioned parameters, as summarized in Table \ref{tab procedure parameters},
will be referred to as procedure parameters.

\section{Examples}

In the following two subsections, we will illustrate our
representation procedure with two slowly diffusing solutions, of
which one diffuses due to a dissipative perturbation, and the
other due to its chaotic nature. In order to illustrate the
convergence property of this procedure, it is convenient to write the resulting
representation with a single term index, i.e.,
\begin{equation}
\label{for v VNAFF} z_v(t) = \sum_{j=1}^J z_j(t) \equiv \sum_{j=1}^J
a_{j}T_{l(j)}(x(t)) \e^{\i\varphi_{\mathbf{k}(j)}(t)}
\end{equation}
where the terms are arranged in the same order as they are obtained in the procedure,
which corresponds roughly to the order of decreasing $|a_j|$.

\figureA
\figureB

\subsection{An example of dissipated solution}

Consider the following weakly dissipated system
\begin{equation}
\label{for D equation} \frac{d^2\theta}{dt^2} = \sin(\theta)
-\varepsilon \frac{d\theta}{dt}
\end{equation}
where
\begin{equation}
\label{for D eqnparameter} \varepsilon=10^{-5}.
\end{equation}
If we nullify the dissipative term $-\varepsilon \frac{d\theta}{dt}$ in
(\ref{for D equation}), the system is a simple pendulum with Hamiltonian
$H_0(I)=\frac{I^2}{2}+\cos(\theta)$, where
$I=\dot{\theta}$.
Take as an example the solution $z(t) \equiv I(t)\e^{\i\
\theta(t)}$ of (\ref{for D equation}) with the following initial
conditions
\begin{equation}
\label{for D initialcondition} t_0=0, \theta_0=0, I_0=1.
\end{equation}
Due to the presence of the dissipative perturbation, the phase orbit starting  from $(\theta_0,I_0)$  decays gradually away from the unperturbed orbit passing through the same phase point. This dissipation effect is significant in the long run, as is shown in Fig.\ref{fig D orbit}.

As a solution of a system of 1 degree of freedom with a dissipative
perturbation, the decaying $z(t)$ has a changing fundamental frequency $\nu$,
which inherits the characteristic frequency of the unperturbed system
and varies as the solution decays.


By numerical integration, an ephemeris of $z(t)$ is obtained at
$\{t_i=ih : i=0,...,65535,h=0.15\}$. We then apply NAFF
to 129 evenly spaced time subintervals with length $d=2046 h$ and midpoints
$\{\tau_{\lambda}=\frac{d}{2}+496(\lambda-1)h\}_{\lambda=1}^{129}$, respectively, to obtain a sample set of the
changing frequency,
$\{\tau_\lambda,\nu(\tau_\lambda)\}_{\lambda=1}^{129}$. This
frequency sample set is shown in Fig.\ref{fig D tvsff}. Also shown is a Chebyshev
expansion approximating $\nu(t)$. This expansion is of degree $9$ and expressed as
\begin{equation}
\nu(t)\approx\sum_{m=0}^{9}c_m T_m(x),
\end{equation}
where the Chebyshev polynomials $T_m(x)$ are given 
in (\ref{for 15 Chebyshev polynomials}), and $x(t)=\frac{2(t-\tau_1)}{\tau_{129}-\tau_1}-1$ 
is the time after being normalized from $[\tau_{1},\tau_{129}]$ to the  $[-1,1]$ interval 
on which the Chebyshev polynomials are defined. 
The values of the coefficients $\{c_{m}\}_{m=0}^{9}$ are listed in Table \ref{tab D FFdata}.
We write the Chebyshev expansion approximating the phase increment, from the value at 
$x=0$ (i.e. $t=\frac{\tau_1+\tau_{129}}{2}$), associated with $\nu$ in two parts
\begin{equation}
\label{phase increment}
\begin{array}{ll}
\int_{(\tau_1+\tau_{129})/2}^t
\mathbf{\nu}(\tau) d\tau &=\frac{\tau_{129}-\tau_1}{2}\int_{0}^{x(t)}
\sum_{m=0}^{9}c_m T_m(x) dx \\[0.5cm]
&=\varphi_0 T_0(x) + \varphi(x).
\end{array}
\end{equation}
In this equation, the time scaling factor $\frac{\tau_{129}-\tau_1}{2}$ accounts for the change of 
integration variable from the physical time to the normalised time, 
 $\varphi_{0}=-\frac{\tau_{129}-\tau_1}{2}\varphi(0)$
 is a real constant, and
\begin{equation}
\label{v phase increment}
\varphi(x)=\frac{\tau_{129}-\tau_1}{2}\sum_{m=1}^{10}C_m T_m(x)
\end{equation}
gathers all Chebyshev polynomials of degree larger than 0
with\footnote{for the general case, see (\ref{for gen IntChbExpansion}).}
 \begin{equation}\label{for IntChbExpansion}
C_m=\left\{
\begin{array}{l}
\frac{2c_{0}-c_{2}}{2}  \hspace{1.3cm} (m=1),\\
\frac{c_{m-1}-c_{m+1}}{2m} \hspace{1cm} (m=2,\ldots ,10).
\end{array}
\right.
\end{equation}
where $c_{10}=c_{11}=0$.

\tableB

\tableC
\figureC
\figureD
\figureE

\figureF
\figureG
\figureH

\figureI
\figureJ
\figureK
\figureL

\tableD
\tableE
\tableF

This completes the determination of $\varphi(x)$, which will be used in the following.
With the procedure parameters
$\{M,K,L,\delta_r\}=\{9,10,9,10^{-5}\}$, we obtain a
37-term representation
\begin{equation}
\label{37 term} z_{37}(t) = \sum_{j=1}^{37} a_{j}T_{l(j)}(x) \e^{\i\varphi_{k(j)}(x)}\ \ \mathrm{with}\ \ x=\frac{2(t-\tau_1)}{\tau_{129}-\tau_1}-1
\end{equation}
of the solution $z(t)$, where $T_{l(j)}(x)$ is the Chebyshev polynomial of degree $l(j)$, and $\varphi_{k(j)}(x)=k(j) \varphi(x)$ the phase increment associated with the frequency $k(j) \nu$. The data needed for specifying the first 10 representation terms are given in Table \ref{tab D termdata}.

Now, we discuss how the procedure parameters affect the precision
of the resulting representation. For this, let's first consider a 150-term representation
obtained by resetting $M=1$. This resetting introduces only a small discrepancy in $\nu(t)$, since  $|c_m/c_0| < 6 \times 10^{-4}$ for
$m>1$. But the precision of this representation is several orders less precise than the previous one, as seen by comparing the two panels of Fig.\ref{fig D precision}.
This is because the whole considered time interval is long, and
so, the small discrepancy in $\nu(t)$ can result in significant
phase errors. On the other hand, however, the fact that there is
no terms with either $|k|>8$ or $l>4$ in the 37-term representation indicates that the representation model practically converges
with respect to $K$ and $L$. Our representation procedure also converges rapidly with respect to $J$. This is illustrated in Fig.\ref{fig D convergence}, for which the procedure parameters excluding $J$ are fixed as $\{M,K,L\}=\{9,10,9\}$.

\subsection{An example of chaotic solutions of Hamiltonian system}

Consider the following Hamiltonian system,
\begin{equation} \label{for H equation}
H(I,\theta,t)=\frac{I^2}{2} + \varepsilon \cos(\theta)[1+\cos(\nu_1 t)+\cos(\nu_2 t)],
\end{equation}
where
\begin{equation} \label{for H eqnparameter}
\varepsilon=5 \times 10^{-3}, \nu_1=1.5,
\nu_2=\frac{\pi}{2}.
\end{equation}
A solution of this system has three fundamental frequencies. The first two
are simply the forcing frequencies $\nu_1$ and $\nu_2$, which are non-commensurable. The
other, denoted as $\nu_3$, can be approximated in the similar way as in the
previous subsection.

By the analysis of the frequency map \citep[e.g.][]{laskar99}, defined as $I_0 \rightarrow \nu_3$ with $\theta_0=0$ and shown in Fig.\ref{fig
CL2 ffmap}, we know that the phase point
\begin{equation}
\label{for H initialcondition} t_0=0, \theta_0=0, I_0=1.535,
\end{equation}
lies in a chaotic zone formed by resonance overlap. And the solution $z(t)$ starting
from this point is chaotic.

An ephemeris of $z(t)$ is obtained by the symplectic integrator
SBABc4 \citep{laskar01} at $\{t_i=ih :
i=0,...,4095,h=0.186058\}$. We then apply NAFF to 129 evenly
spaced time intervals with length $d=512 h$ and midpoints
$\{\tau_{\lambda}=\frac{d}{2}+28(\lambda-1)h\}_{\lambda=1}^{129}$, respectively. This gives the sample set
$\{\tau_{\lambda},\nu_{3,\lambda}\}_{\lambda=1}^{129}$, partly
shown in Fig.\ref{fig CL2 tvsff} together with a Chebyshev approximation. It should be noted that there
are two intrinsically different error sources in the present way
of approximating a changing frequency. One is related to the NAFF
process that gives the samples of the frequency, while the other
related to the fitting process that leads to a Chebyshev approximation of
the frequency. Accordingly, we consider the Chebyshev approximation
as sufficiently good if it deviates from the frequency
sample set much less than the sample uncertainties. As shown in
Fig.\ref{fig CL2 tvsff}, this requirement can be met with the
Chebyshev polynomial of degree $M_3=9$.

Test calculations show that our representation procedure cannot lead to an
acceptable representation of the solution on the whole sampling time interval.
There are two possible reasons for this.

The most intrinsic reason would be
that our solution experiences passages into or out of resonance zones. Such a
passage is associated with a shift between circulation and libration of the
corresponding resonance angle, and so, with occurrence or disappearance of
certain terms. If some of these terms are significant enough in the whole considered time interval $(\tau_1,\tau_{129})$, then there would be no way to get any acceptable non-piecewise representation. Therefore, we will restrict ourselves to the shorter time interval
$(\tau_1,\tau_{50})$.

Another possible reason is that there is one or more
significant libration frequencies, which are not taken into consideration when
we generate the function basis $B$. While it is easy to make a necessary
extension of $B$ in order to include known libration frequencies (see section \ref{sec 5}), it is
not that straightforward to identify and sample these frequencies
\citep{laskar90}. To show the flexibility of our algorithm, we will not search
for any libration frequency and make the corresponding extension of $B$. The
flexibility comes from the fact that, if we choose reasonably large values of
our procedure parameters $K_n$'s, then the resulting set of frequencies would be dense enough over a large frequency interval, in the sense
that every important libration frequency is not far from at least one element of the frequency set.

Setting the procedure parameters as $$\{M_1,M_2,M_3,K_1,K_2,K_3,L_\mathbf{k},J\}=\{0,0,9,10,10,10,9,100\},$$ we
obtain a 100-term representation of our solution. Fig.\ref{fig CL2 precision}
shows that the errors of this representation are typically of order less than
$10^{-4}I_0$. And, Fig.\ref{fig CL2 convergence}
illustrates, in the same way as in the previous subsection, the
convergence property of the representation procedure.

\section{Application to planetary ephemerides  \label{sec 5}}

Numerical integration is now an efficient way of constructing ephemerides of
the solar system bodies  with high precision. For practical applications, however, it
can be  useful to represent analytically, and thus in a continuous way, these discrete solutions.

These representations can be done in the form of segmented Chebyshev
expansions, like the ones representing the classical planetary ephemerides as INPOP
\citep[e.g.][]{fienga08}, or other generally applied approximation models without physical basis.
A drawback of these  representations is that they
require large amount of data. In order to get compact representations,
\citet{chapront} uses a model of Poisson series, with fixed main
frequencies, that were   obtained with the  NAFF algorithm. Though this model already
involves some long-term or long-period-term effects by allowing Poisson terms, and as thus, works
well with planetary ephemerides of five outer planets over a few
hundred years, it does not take into consideration the frequency drifts, and cannot be used over million of years.

The frequency drifts are important over a few tens of million years, as shown by
\citet{laskar90}. The algorithm developed in the present paper should thus be more
appropriate in representing ephemerides spanning this long time interval. It is
thus interesting to test whether the present algorithm can be used to represent over such
a time scale the long-term numerical solution of major solar system bodies,
\citep[e.g.][]{laskar04}. For this, we apply our algorithm to the eccentricity and inclination variables of the Earth, i.e.,
\begin{equation} \label{for z3zeta3}
z_3 =e_3\exp(i\varpi_3)\ \ \mathrm{and} \ \
\zeta_3=\sin(i_3/2)\exp(i\Omega_3).
\end{equation}
To be more precise, $e_3\ \mathrm{and}\ i_3$ are the eccentricity and inclination, respectively, of the instantaneous orbit of the Earth-Moon barycenter, and
$\varpi_3\ \mathrm{and}\ \Omega_3$ are, respectively, the longitudes of the
perihelion and of the node of the same orbit with respect to the fixed J2000.0
equatorial reference system.

The chaotic behaviour of the terrestrial orbit certainly limit the time span over
which this orbit can be precisely determined, but $z_3$ and $\zeta_3$ from
La2004 are reliable and precise at least over the time span $[-35 \mathrm{Myr},+5
\mathrm{Myr}]$  \citep{laskar11}. Therefore, we restrict our
representations to this time span. Test calculations show that our algorithm can
lead to compact and precise representations for both degrees of freedom.
With different procedure parameters, the resulting representations contain very
different terms. This confirms the flexibility of the present algorithm.

In order to give representations as physical as possible, we will resort to the knowledge
we have for the solution. Following \citet{laskar04}, we compute the fundamental frequencies of the secular solar system
by applying the NAFF algorithm over time intervals of length 20 Myr for the proper modes
$(z^\bullet_1,...,z^\bullet_4,\zeta^\bullet_1,...,\zeta^\bullet_4)$, and 50 Myr for the proper modes
$(z^\bullet_5,...,z^\bullet_8,\zeta^\bullet_5,...,\zeta^\bullet_8)$, respectively\footnote{
See \citep{laskar90} for the definition of the proper modes.}. From -35 Myr to 5 Myr
with step 0.1 Myr, we generate the samples of the fundamental frequencies $(g_1,...,g_8,s_1,...,s_8)$ corresponding
to these proper modes. The resulting nominal value of $s_5$ is about  $0.00000015\ \mathrm{arcsec\ yr}^{-1}$.
The other frequency samples are shown respectively in the panels of Fig.\ref{fig solar FFs}, where the errors are
estimated as the difference between the values of a frequency computed from the associated ephemeris and its quasiperiodic
approximation. Also shown in this figure are the Chebyshev approximations of these
fundamental frequencies. All of these Chebyshev approximations, the coefficients of which are listed in Table \ref{tab FFChC},
are obtained respectively by truncating the ones of degree 15. The truncation criterion is roughly that the discrepancy
in a fundamental frequency should not induce an error in phase larger than $2\pi/10^4$ over several tens of million years.

There are two important libration frequencies, i.e. $r_1=0.251085\ \mathrm{arcsec\ yr}^{-1}$ of the resonance argument
$2(\varpi^\bullet_4-\varpi^\bullet_3)-(\Omega^\bullet_4-\Omega^\bullet_3)$ and $r_2=0.117222\ \mathrm{arcsec\ yr}^{-1}$
of $2(\varpi^\bullet_1-\varpi^\bullet_5)-(\Omega^\bullet_1-\Omega^\bullet_2)$. To include them as additional fundamental
frequencies, we express a main frequency as
\begin{equation} \label{MainFrequency}
\omega=\sum_{i=1}^8 (m_ig_i+n_is_i)+\sum_{j=1}^2k_j r_j,
\end{equation}
The d'Alembert characteristic, i.e. $\sum_{i=1}^8
(m_i+n_i)=1$, will be used to exclude non-physical frequency index vectors.

To show to what degree our algorithm is practically useful, we discuss here
the following two 100-term representations ($\Re_{100}$ for short),
\begin{equation}
\label{z zata 100} z_3(t)=\sum_{j=1}^{100} a_{\mathbf{k}(j)} \e^{\i \varphi_{\mathbf{k}(j)}(t)},\ \
 \zeta_3(t)=\sum_{j=1}^{100} b_{\mathbf{k}(j)} \e^{\i \varphi_{\mathbf{k}(j)}(t)},
\end{equation}
where, with $\mathbf{f}=(g_1,...,g_8,s_1,...,s_8,r_1,r_2)$ and $\mathbf{k}=(m_1,...,m_8,n_1,...,n_8,k_1,k_2)$,
the phase increment $\varphi_\mathbf{k}(t)=\int_0^t \langle \mathbf{k},\mathbf{f(\tau)} \rangle d\tau$ is associated
with the main frequency $\langle \mathbf{k},\mathbf{f} \rangle$. The way the residuals decrease with the increasing
number of representation terms in the two representation procedures of $\Re_{100}$ are shown in Fig.\ref{fig convergencela04}.

The leading 40 terms of these two representations are given, respectively,
in Tables \ref{tab MainFrequencyz3} and \ref{tab MainFrequencyzeta3}, where the terms are reordered according
to their real amplitudes and data are rounded to a convenient number of digits. Comparing these results with
the ones given in tables 4 and 5 of \citet{laskar90}, we find that they are coherent with each other
in the sense that all the main frequencies explicitly identified previously can be found in the present tables.

The on-line electronic files associated with the present paper, i.e. z3R100.dat and zeta3R100.dat, are the full version of Table \ref{tab MainFrequencyz3} and Table \ref{tab MainFrequencyzeta3}, respectively. They are plain text tables providing all terms of $\Re_{100}$  in the form of (\ref{z zata 100}). Together with the data presented in Table \ref{tab FFChC} for computing $\mathbf{f}$, these two tables can be used to compute the eccentricity and inclination variables from $\Re_{100}$. The errors of $\Re_{100}$ as solution representation are shown in Fig.\ref{fig EphvsRepla04}. From this figure, we expect that the present algorithm should be efficient in producing compact and precise representations of long-term ephemerides of major solar system bodies.

To conclude, we illustrate explicitly the advantage in the representation of ephemerides of taking into consideration the frequency
drifts by comparing the representations given respectively by a direct use of NAFF and the modified present algorithms.
The quasi-periodic representations ($\Re_{NAFF}$ for short) of the same
$z_3$- and $\zeta_3$-ephemeris given by the standard realization of NAFF, which is a
built-in tool of the algebraic system TRIP (http://www.imcce.fr/trip/), have respectively 50 and 41 terms.
NAFF stops recovering more terms because it encounters a frequency that, at a
given level of precision, is already recovered in a previous step. To show more
precisely the advantage of taking frequency drifts into
consideration, we produce for $z_3$ and
$\zeta_3$, respectively, a representation ($\Re$) with the same number of terms as the corresponding $\Re_{NAFF}$ representation.
The comparison between $\Re$ and $\Re_{NAFF}$ is shown in Fig.\ref{fig
VqpQpla04}. From this figure, we see clearly the improvements brought by
introducing frequency drifts into the representation model.

\begin{acknowledgements}
P. Robutel and L. Niederman are thanked for their time in
instructive discussions, and M. Gastineau for various kinds of
helps. Fu is indebted to many colleagues from IMCCE
for their hospitality. Fu is supported by IMCCE and NSFC under Grant Nos. 11178006 and 11533004.

\end{acknowledgements}

\bibliographystyle{aa}  
\bibliography{naffdif} 

%

\begin{appendix}
\onecolumn
\section{Chebyshev polynomials}

Chebyshev polynomials as defined by the following recurrence relation
\begin{equation}\label{for Chebyshev polynomials}
\begin{array}{l}
 T_0(x)=1,\ \ T_1(x)= x \hspace{3.75cm} (-1 \leq x \leq 1)  \\
 T_{m+1}(x)=2xT_m(x)-T_{m-1}(x)\hspace{2.8cm}(m > 1)
\end{array}
\end{equation}
form a non-normalized but orthogonal basis under the inner product
\begin{equation}\label{for ChbOrtInnerproduct}
\langle f(x),g(x)\rangle=\int_{-1}^{1}\frac{f(x)g(x)}{\sqrt{1-x^2}}dx.
\end{equation}
This can be easily checked by straightforward calculations
\begin{equation}\label{for ChbOrtnonNor}
\langle T_i(x),T_j(x)\rangle=\int_{-1}^{1}\frac{T_i(x)T_j(x)}{\sqrt{1-x^2}}dx=
\left\{
\begin{array}{l}
\pi\hspace{1.28cm}  (i=j=0) \\
\pi/2\hspace{1cm}  (i=j\neq 0)\\
0\hspace{1.31cm}(i\neq j).
\end{array}
\right.
\end{equation}

Their linear combination, called Chebyshev expansion, is often used to approximate a function $h(x)$ defined on $[-1,1]$
\begin{equation}\label{for ChbExpansion}
h(x)\approx h_c(x)=\sum_{m=0}^{M}c_m{T_m(x)},
\end{equation}
where
\begin{equation}\label{for ChbExpansion1}
c_m=\left\{
\begin{array}{l}
\frac{1}{\pi}\langle h(x),T_m(x)\rangle\hspace{1.28cm}  (m=0) \\ \\
\frac{2}{\pi}\langle h(x),T_m(x)\rangle\hspace{1.28cm}  (m\neq 0).
\end{array}
\right.
\end{equation}

The indefinite integral of the Chebyshev expansion $h_c(x)$ writes, up to an arbitrary constant,
\begin{equation}\label{for ChbExpansion}
H_c(x)=\sum_{m=1}^{M+1}C_m{T_m(x)},
\end{equation}
where, with $c_{M+1}=c_{M+2}=0$,
\begin{equation}\label{for gen IntChbExpansion}
C_m=\left\{
\begin{array}{l}
\frac{2c_{0}-c_{2}}{2} \hspace{1.3cm} (m=1),\\
\frac{c_{m-1}-c_{m+1}}{2m} \hspace{1cm} (m=2,\ldots,M+1).
\end{array}
\right.
\end{equation}

The explicit expressions of the Chebyshev polynomials up to degree 15 are listed below
\begin{equation}\label{for 15 Chebyshev polynomials}
\renewcommand{\arraystretch}{1.5}
\begin{array}{l}
 T_0(x)=1 \\
 T_1(x)= x \\
 T_2(x)=2 x^2-1 \\
 T_3(x)=4 x^3-3 x \\
 T_4(x)=8 x^4-8 x^2+1 \\
 T_5(x)=16 x^5-20 x^3+5 x \\
 T_6(x)=32 x^6-48 x^4+18 x^2-1 \\
 T_7(x)=64 x^7-112 x^5+56 x^3-7 x \\
 T_8(x)=128 x^8-256 x^6+160 x^4-32 x^2+1 \\
 T_9(x)=256 x^9-576 x^7+432 x^5-120 x^3+9 x \\
 T_{10}(x)=512 x^{10}-1280 x^8+1120 x^6-400 x^4+50 x^2-1 \\
 T_{11}(x)=1024 x^{11}-2816 x^9+2816 x^7-1232 x^5+220 x^3-11
   x \\
 T_{12}(x)=2048 x^{12}-6144 x^{10}+6912 x^8-3584 x^6+840
   x^4-72 x^2+1 \\
 T_{13}(x)=4096 x^{13}-13312 x^{11}+16640 x^9-9984 x^7+2912
   x^5-364 x^3+13 x \\
 T_{14}(x)=8192 x^{14}-28672 x^{12}+39424 x^{10}-26880
   x^8+9408 x^6-1568 x^4+98 x^2-1 \\
 T_{15}(x)=16384 x^{15}-61440 x^{13}+92160 x^{11}-70400
   x^9+28800 x^7-6048 x^5+560 x^3-15 x.
\end{array}
\renewcommand{\arraystretch}{1.5}
\end{equation}

\end{appendix}
\end{document}